%% file: IN21_novel_features_bearing_fault.tex
\documentclass[a4paper,12pt]{article}

\usepackage[utf8x]{inputenc}
\usepackage[T1]{fontenc}
\usepackage{times}
\usepackage[colorlinks=true,linkcolor=black,citecolor=black,urlcolor=blue]{hyperref}
\usepackage{geometry}
\usepackage{fancyhdr}

\usepackage{color}
\usepackage{tikz}
\usepackage{pgfplots}
\usetikzlibrary{calc}
\usepackage{subfigure}
\usepackage{caption}
\usepackage[nolist,nohyperlinks]{acronym}
\usepackage{titlesec}
\titleformat{\section}
{\bfseries\uppercase}{\thesection.}{1em}{}
\titleformat{\subsection}
{\bfseries}{\thesubsection.}{1em}{}

\usepackage{graphicx} 
\usepackage{multirow} 
\usepackage{cite}
\usepackage{breakurl}
\usepackage{indentfirst}
\usepackage{amsmath, amssymb, amsfonts, bm}
\usepackage{cleveref} 

\usepackage{txfonts}
\usepackage{enumitem}
\usepackage{xcolor}
\usepackage{enumitem}
\titlespacing*{\subsection}{0pt}{1.5em}{0.2em}

\newlength{\bibitemsep}
\newlength{\bibparskip}
\let\oldthebibliography\thebibliography
\renewcommand\thebibliography[1]{%
  \oldthebibliography{#1}%
  \setlength{\parskip}{\bibitemsep}%
  \setlength{\itemsep}{\bibparskip}%
}
\newlength\fheight 
\newlength\fwidth 
\begin{document}

\begin{acronym}
\acro{AMS}{Amplitude Modulation Spectrogram}
\acro{BA}{Balanced Accuracy}
\acrodefplural{BA}{Balanced Accuracies}
\acro{DCT}{Discrete Cosine Transform}
\acro{FN}{False Negative}
\acro{FNR}{False Negative Rate}
\acro{FP}{False Positive}
\acro{FPR}{False Positive Rate}
\acro{DFT}{Discrete Fourier Transform}
\acro{DTFT}{Discrete-Time Fourier Transform}
\acro{IDFT}{Inverse Discrete Fourier Transform}
\acro{MFCC}{Mel Frequency Cepstral Coefficient}
\acro{STFT}{Short-Time Fourier Transform}
\acro{SVM}{Support Vector Machine}
\acro{RMS}{Root Mean Square value}
\acro{rpm}{Rotations per minute}
\acro{TN}{True Negative}
\acro{TNR}{True Negative Rate}
\acro{TP}{True Positive}
\acro{TPR}{True Positive Rate}
\end{acronym}

\begin{center}
	\includegraphics[width=38.8mm, height=20.6mm]{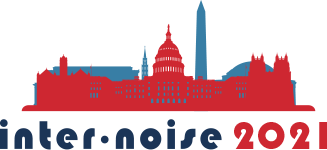}
\end{center}
\vskip.5cm

\begin{flushleft}
\fontsize{16}{20}\selectfont\bfseries
Novel features for the detection of bearing faults in railway vehicles
\end{flushleft}
\vskip1cm

\renewcommand\baselinestretch{1}
\begin{flushleft}

Matthias Kreuzer \footnote{matthias.kreuzer@fau.de}\\
Multimedia Communications and Signal Processing \\
Friedrich-Alexander University Erlangen-Nürnberg \\
Cauerstr. 7, 91058 Erlangen, Germany \\
\vskip.5cm
Alexander Schmidt\footnote{alexander.as.schmidt@fau.de}\\
Multimedia Communications and Signal Processing \\
Friedrich-Alexander University Erlangen-Nürnberg \\
Cauerstr. 7, 91058 Erlangen, Germany \\
\vskip.5cm
Walter Kellermann\footnote{walter.kellermann@fau.de}\\
Multimedia Communications and Signal Processing \\
Friedrich-Alexander University Erlangen-Nürnberg \\
Cauerstr. 7, 91058 Erlangen, Germany \\

\end{flushleft}

\textbf{\centerline{ABSTRACT}}\\
\textit{In this paper, we address the challenging problem of detecting bearing faults from vibration signals. For this, several time- and frequency-domain features have been proposed already in the past. However, these features are usually evaluated on data originating from relatively simple scenarios and a significant performance loss can be observed if more realistic scenarios are considered. To overcome this, we introduce \acp{MFCC} and features extracted from the \ac{AMS} as features for the detection of bearing faults. Both AMS and MFCCs were originally introduced in the context of audio signal processing but it is demonstrated that a significantly improved classification performance can be obtained by using these features. Furthermore, to tackle the characteristic data imbalance problem in the context of bearing fault detection, i.e., typically much more data from healthy bearings than from damaged bearings is available, we propose to train a  One-class \ac{SVM} with data from healthy bearings only. Bearing faults are then classified by the detection of outliers. Our approach is evaluated with data measured in a highly challenging scenario comprising a state-of-the-art commuter railway engine which is supplied by an industrial power converter and coupled to a load machine. \\
}
\section{INTRODUCTION}
\noindent
Bearings represent crucial components in rotating machinery, e.g., wind turbines and railway vehicles. As a 
bearing failure may cause substantial additional costs and also damages to other components, an early and reliable detection of bearing faults is of major importance. Since these components are often difficult to access, manual inspection is time-consuming and expensive. However, damages to the bearing alter the vibration signature of the machine, which can be captured by acceleration sensors that are installed on the housing of the machine. Consequently, faults can be identified by analyzing the vibration signal.

For this, various techniques have been developed over the recent decades. The analysis of the vibration signals is typically performed in the time domain or the frequency domain. In this context, especially the so-called fault frequencies play a crucial role: If one of the four major bearing components, i.e., the inner race, the outer race, the cage and the rolling elements, is damaged, impulses with component-specific frequencies can be observed in the vibration signal.  
Many methods analyze the spectrum of the envelope of the vibration signal  as the fault frequencies become more apparent after demodulating the signal \cite{randall_rb_rolling_2011}. Yet, the signal components that are related to the bearing fault are often buried in environmental noise caused by other machine components, which necessitates the application of signal enhancement algorithms. In order to separate the fault-related components from the environmental noise, several signal decomposition techniques, e.g., the Empirical Mode Decomposition \cite{zhang2017a} and  Wavelet Transforms \cite{wei2019}, and adaptive filtering techniques \cite{Elasha2014} have been applied to this problem. Detailed reviews for signal processing techniques applied to the topic of bearing fault detection can be found in \cite{wei2019}. 

As the vibration pattern changes with the appearance of a fault, statistical features, e.g., the variance, the kurtosis etc., of the vibration signal also change which can be be used to detect a fault. Consequently, data-driven machine learning classification algorithms, e.g., linear discriminant analysis, k-nearest neighbors, \acp{SVM} and neural networks, can be trained on features extracted from the time- and frequency-domain representations of the vibration signal \cite{hamadache2019}. Naturally, the performance of these classifiers is strongly dependent on the extracted features. Thus, determining the best-suited features is crucial for this kind of approach.
Recently, also end-to-end deep learning-based approaches have gained wide popularity \cite{neupane2020}. In contrast to the previous classification algorithms, a manual feature selection step is not required as the features are extracted by the network itself. A comprehensive overview for a wide variety of deep learning architectures including autoencoders, convolutive neural networks and recurrent neural networks can be found in \cite{neupane2020}. 
Although the manual feature selection step can be omitted, most of these approaches suffer from the fact that they require a large amount of labeled training data for both healthy and damaged bearings in order to show satisfactory results. However, the available data is often unbalanced as data for damaged bearings in realistic scenarios is hard to obtain, e.g, due to safety reasons, whereas data for healthy bearings is usually available in abundance. Furthermore, the labeling process for naturally developed damages is difficult as it is hard to determine precisely the point in time when the first signs of a fault begin to show. Hence, approaches that are not reliant on data from damaged bearings or only require a very little amount, are especially well-suited for the classification of bearing faults. This type of classification problem is referred to as outlier or anomaly detection in the literature. 

In this paper, we evaluate and compare various features from the time and frequency domain for the detection of bearing faults in a highly challenging scenario. The experimental setup consists of a state-of-the-art commuter railway engine coupled to a load machine. Both engine and load are supplied by an industrial power converter. In contrast to most of the publicly available datasets the rotational frequency and the applied torque vary strongly and the power converter strongly affects the measured vibration patterns.
In addition to the well-studied features from literature, we also investigate novel features from the domain of audio signal processing and automatic speech recognition (ASR), namely \acf{AMS} and \acfp{MFCC}. It is shown that these features are highly effective and outperform conventional time- and frequency-domain features. Further, in order to address the commonly encountered data imbalance problem we only use features obtained from healthy bearings to train a One-Class \acf{SVM} and treat the classification of bearing faults as an anomaly detection task. It is shown that even with this rather simple classifier and an appropriate set of features, classification accuracies over 98~$\%$ can be achieved.

This article is organized as follows. In  Section~\ref{sec:signal_model}, the underlying signal model is introduced before conventional features for bearing fault detection from time and frequency domain are briefly reviewed in Section~\ref{sec:features}. After this, the \acp{MFCC} and the \ac{AMS}  are discussed in more detail in Section~\ref{sec:mfcc} and Section~\ref{sec:ams}, respectively. In Section~\ref{sec:svm}, the One-Class \ac{SVM} classification method is described.  In Section~\ref{sec:evaluation}, the experimental setup (cf. Section~\ref{sec:experimental_setup}), the data generation and the training process along with used evaluation metrics (cf. Section~\ref{sec:data}) are described, before the experimental results are discussed in detail in Section~\ref{sec:results}. A  summary in Section~\ref{sec:summary} concludes the paper.

\section{Bearing Fault Classification}
\label{sec:bearing_fault_classification}
\noindent
In this section, the investigated features, most importantly the proposed \ac{AMS} and \acp{MFCC}, are introduced and the classification method is described. 
\subsection{Signal Model} \label{sec:signal_model}

\noindent
In the following, the discrete time-domain vibration signal will be denoted as $x[k]$ with sample index $k$, which is obtained after sampling the continuous-time domain vibration signal with sampling frequency $f_s$. 
Given $K$ samples of the discrete-time signal $x[k]$, the \ac{DFT} can be defined as follows  
\begin{align}
    X[\mu] = \mathrm{DFT}_{M} \{x[k] \} = \sum_{k=0}^{K-1} x[k] e^{-j\frac{2 \pi}{M} k \mu },
\label{eq:dft}
\end{align}
where $\mu$ and $M$ denote the frequency bin index and the length of the \ac{DFT}, respectively. 
The \ac{STFT} of $x[k]$ for a segment of length $K$  is then given by
\begin{equation}
    X[\mu, n] = \sum_{k=n}^{n+K-1} x[k] w_M[k-n] e^{-j\frac{2 \pi}{M} k \mu},
    \label{eq:stft}
\end{equation}
with time frame index $n$. $w_M[k]$ is a window function of length $M$ \cite{oppenheim}.

\subsection{Conventional Time- and Frequency-domain Features}
\label{sec:features}
\noindent
In this section, we briefly review conventional features from the time domain and the frequency domain. Subsequently, the proposed features, i.e., the  \acp{MFCC} and \ac{AMS}, are introduced in Sections~\ref{sec:mfcc} and \ref{sec:ams}, respectively. The selected features from the time domain include  average, variance, root mean square value (RMS), kurtosis, skewness, amplitude range and peak-to-RMS ratio of $x[k]$ and according estimators are summarized in Table~\ref{Tab:1}. These features are well-established for the detection for bearing faults and used for example used in \cite{vakharia_comparison_2016, vargas_2020, nayana_analysis_2017,br_feature_2019}. In the following, the set of time-domain features is abbreviated with \textit{TD}. Analogously, statistical features can be derived from the frequency-domain vibration signal $X[\mu]$. We consider spectral centroid, spectral spread, spectral kurtosis,  spectral entropy,  spectral crest and roll-off point (cf. Table~\ref{Tab:2}). These features have been successfully applied for bearing fault detection in \cite{arun2018,yuan_fault_2020,vargas_2020}. Henceforth, the set of features listed in Table~\ref{Tab:2} is denoted as \textit{SD}, i.e., the spectral descriptors. 
As localized faults can be linked to characteristic frequencies, the amplitudes at multiples of the characteristic fault frequencies in the envelope magnitude spectrum are also considered as features. These features are computed as
\begin{align}
            AMP_{fault} = \sum_{i=1}^3 |X_{env}[i \cdot \mu_{fault}]|  \label{eq:env_amp} \\
            \text{with} \quad \mu_{fault} \in \left\{ \mu_{\text{BPFO}}, \mu_{\text{BPFI}}, \mu_{\text{CA}}, \mu_{\text{RE}} \right\}, \nonumber 
\end{align}
where $X_{env}[\mu]$ denotes the envelope spectrum (cf. \cite{randall_rb_rolling_2011}), which is obtained by applying the Hilbert transform \cite{oppenheim} and $\mu_{fault}$ is the frequency bin index corresponding to the characteristic fault frequencies $f_{\text{BPFO}}$, $f_{\text{BPFI}}$, $f_{\text{CA}}$ and $f_{\text{RE}}$ for a localized fault in the outer race, the inner race, the rolling cage and the rolling elements, respectively. Equations to determine the characteristic fault frequencies can be found in \cite{randall_rb_rolling_2011,zhang2017a}.

\begin{table}[tbh]

\caption{Overview of considered time-domain features. }
\label{Tab:1}

\begin{center}
\begin{tabular}{l l} 
 \hline 
 \textbf{Feature} &  \textbf{Formula} \\ [0.5ex] 
 Average $ \bar{x}$ & \large $  \frac{1}{K-1}\sum_{k=0}^{K-1} x[k]$ \\ [1ex]   
 
 \hline \\ [-2ex]
Variance $\sigma^2_{x}$  & \large $ \frac{1}{K-1}\sum_{k=0}^{K-1} \left( x[k] - \bar{x} \right)^2$  \\ [1ex] 
 \hline \\ [-2ex]
 Root Mean Square (RMS) $x_{\mathrm{RMS}}$ & \large  $ \sqrt{\frac{1}{K-1}\sum_{k}^{K} x[k]^2} $ \\ [1ex] 
\hline \\ [-2ex]
Kurtosis & \Large $\frac{\frac{1}{K-1}\sum_{k=0}^{K-1} \left( x[k]-\bar{x}\right) ^4}{(\sigma^2_{x})^2}$ \\ [1ex] 
\hline \\ [-2ex]
Skewness & \Large $   \frac{\frac{1}{K-1} \sum_{k=0}^{K-1} \left( x[k]-\bar{x}\right) ^3}{\left(\sqrt{\sigma^2_{x}}\right)^3}$ \\ [2ex] 
\hline \\[-2ex]
 Amplitude Range & \large $ \max(x[k]) -\min(x[k])$\\  
\hline \\ [-2ex] 
Peak-to-RMS ratio  & \Large $ \frac{  \max(x[k]) }{x_{\mathrm{RMS}}}  $\\
 [1ex] 
 \hline
\end{tabular}
\end{center}
\end{table}

\begin{table}[tbh]
\captionsetup{type=table,format=hang}
\caption{Overview of considered frequency-domain features. $\mu_1=0$ and $\mu_2=K-1$ denote the lower and upper frequency bin index, respectively, $f_\mu$ denotes the frequency corresponding to frequency bin $\mu$. $\kappa$ is set to 0.95.}
\label{Tab:2}
\begin{center}
\begin{tabular}{l l } 
\hline
 \textbf{Feature} &  \textbf{Formula}  \\ [0.5ex] 
 \hline \\
Spectral Centroid (SC) & \large   $ \frac{\sum_{\mu=\mu_1}^{\mu_2} f_\mu  |X[\mu]|}{\sum_{\mu=\mu_1}^{\mu_2} |X[\mu]|} $  \\ [3ex]
 \hline \\ [-2ex]
 Spectral Spread (SSpr) & \large   $ \sqrt{\frac{\sum_{\mu=\mu_1}^{\mu_2} (f_\mu -\mathrm{SC} )^2 | X[\mu] |}{ \sum_{\mu=\mu_1}^{\mu_2} | X[\mu] |}}$  \\  [3ex] 
 \hline \\ [-2ex]
 Spectral Kurtosis & \large  
        $\frac{\sum_{\mu=\mu_1}^{\mu_2} (f_\mu  - \mathrm{SC})^4 |X [\mu]|}{(\mathrm{SSpr})^4 \sum_{\mu=\mu_1}^{\mu_2} |X[\mu]|}$  \\  [1ex] 
    \hline \\  [-2ex]
    Spectral Entropy & \large  $ \frac{-{\sum_{\mu=\mu_1}^{\mu_2} |X[\mu]| \log(|X [\mu]|)} }{\log(\mu_2-\mu_1)}$ \\  [1ex] 
    \hline \\ [-2ex]
    Spectral Crest & \large  $\frac{\max(|X[\mu]|)} {\frac{1}{\mu_2-\mu_1 +1}\sum_{\mu=\mu_1}^{\mu_2} |X[\mu]|}$ \\  [2ex] 
    \hline \\ [-2ex]
    Roll-off point & \normalsize  $ \sum_{\mu=\mu_1}^{i} |X[\mu]| = \kappa \sum_{\mu=\mu_1}^{\mu_2} |X[\mu]|$  \\  [1ex] 
    \hline 
\end{tabular}
\end{center}
\end{table}
\subsection{Mel Frequency Cepstral Coefficients}
\label{sec:mfcc}
\noindent
\acp{MFCC} are state-of-the-art features for automatic speech recognition tasks and acoustic scene classification. As the name suggests, the MFCCs combine the cepstrum  with a Mel-frequency scaling which allows for a compact representation of the frequency spectrum of a signal. In speech analysis, the real-valued cepstrum, i.e., the \ac{IDFT} of the log magnitude of the \ac{DTFT} of a signal, is widely popular for various applications such as, e.g., pitch detection and acoustic scene classification \cite{schafer2007}. 
The \ac{MFCC} feature extraction scheme can be summarized as follows:
\begin{enumerate}
    \item The time domain signal is divided into frames by windowing.
    \item An $N$-point \ac{DFT} is computed of the windowed input samples.
    \item The resulting power magnitude spectrum is then filtered using a Mel-filterbank, which is approximated, e.g., by applying triangular windows in the \ac{DFT} domain \cite{schafer2007}.
    \item Finally, the \acp{MFCC} are obtained by computing the \ac{DCT} of the logarithm of summed filter bank energies.
\end{enumerate}
For time-frame $n$, the spectral energies $X_{i}[n]$ of time frame $i$ with $i =0,\ldots, K-1$ can be computed as 
\begin{align}
    X_i[n] = \sum_{\nu=0}^{N-1} g_{i\nu} \left| \sum_{k=0}^{N-1}  x[n-k] w[k] e^{-\frac{2\pi k\nu}{N}}\right|^2,
\end{align} 
where $x[k]$ are in the input samples, $w[k]$ is a window function and $g_{i\nu} $ is a triangular window function of a Mel-filterbank.
The Mel filters mimic the human auditory system which has a finer resolution for lower frequency regions than for higher frequency regions. The bandwidths of the practically used Mel filters are constant for center frequencies below $500$~Hz and increase approximately exponentially for higher center frequencies. The number of subbands, i.e., number of filters, varies typically from $K=24$ to $K=40$ subbands.
After filtering, the \acp{MFCC} $c[{\mu}]$ are obtained after applying a \ac{DCT}
\begin{align}
    c[\mu] = \sum_{i=1}^{K} \log X_{i}[n] \cos\left( \frac{\pi (2i-1)\mu} {2K} \right), \; \mu = 1, \ldots , K.
\end{align}
which remove the correlation between the cepstral coefficients introduced by the overlapping filterbanks.

\subsection{Amplitude Modulation Spectrogram}
\label{sec:ams}
\noindent
The \ac{AMS} is a popular feature in hearing-related signal processing and speech recognition and is known for its usefulness  for the detection of envelope fluctuations of speech signals \cite{Kollmeier1994}.
\begin{figure}[t]
    \centering
    \input{figures/ams.tex}
    \caption{Block diagram for AMS.}
\end{figure}
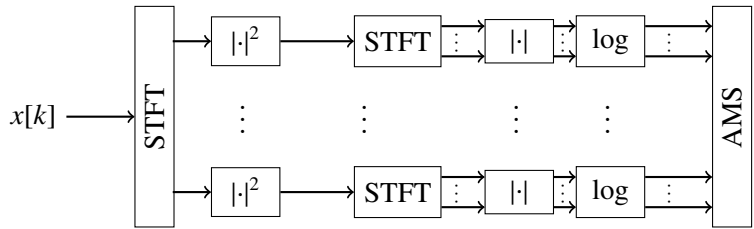
\label{fig:ams}
The signal processing steps that are required for computing the \ac{AMS}
are illustrated in the block diagram  in Figure \ref{fig:ams}. First, the time-domain signal $x[k]$ is transformed to the STFT domain. In the next step, a second \ac{STFT} of the squared magnitude of the complex-valued \ac{STFT} coefficients for every frequency subband is computed. The  subband signals after the squaring operation can be interpreted as the envelope signal of the spectral amplitudes for every subband. This
second \ac{STFT} can be understood as computing a Fourier series of the temporal envelope signal for a given frequency bin, and thereby detecting periodic components (modulation frequencies) in the given frequency subband.
Finally, the \ac{AMS} is obtained by compressing the magnitude logarithmically.
Sometimes the number of subbands is reduced before the second \ac{STFT} by applying a Mel- or Bark-filterbank \cite{Kollmeier1994, Tchorz2018}. In this article the filtering step is omitted and the parameters listed in Table~\ref{Tab:param_nfft} are utilized for the computation of the two consecutive \acp{STFT}. 

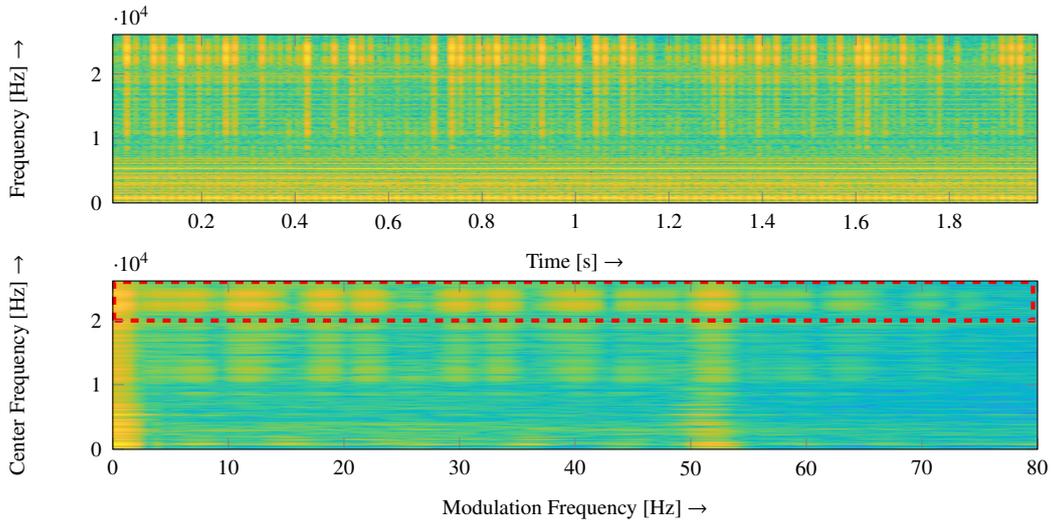
\begin{figure}[h]
        \setlength{\fwidth}{0.85\textwidth}
        \setlength{\fheight}{5.5cm}
        \centering
        \input{figures/ams_damaged.tex}
        \caption{Spectrogram (top) and AMS (bottom) for a damaged bearing with a fault in the outer race.}
        \label{fig:ams_damaged}
\end{figure}

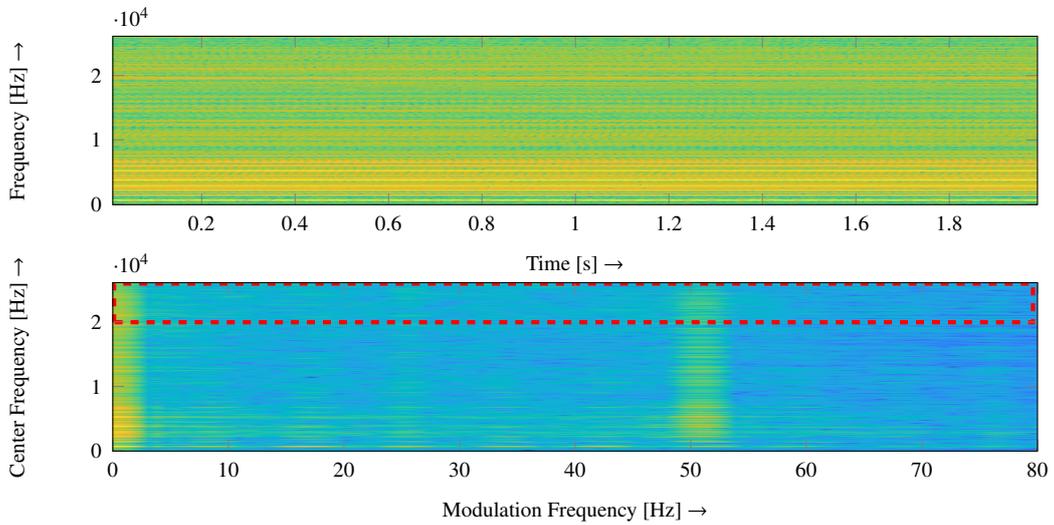
\begin{figure}[h]
        \setlength{\fwidth}{0.85\textwidth}
        \setlength{\fheight}{5.5cm}
        \centering
        \input{figures/ams_healthy.tex}
        \caption{Spectrogram (top) and AMS (bottom) for a healthy bearing.}
        \label{fig:ams_healthy}
\end{figure}

To illustrate that the \ac{AMS} can be a valuable feature for the detection of bearing faults, we consider Figures~\ref{fig:ams_damaged} and \ref{fig:ams_healthy}. Figure~\ref{fig:ams_damaged} shows the spectrogram (top) and the \ac{AMS} (bottom) for a vibration signal excerpt of $2$~s for a bearing with a localized fault in the outer race and a rotational frequency of $500$~rpm ($\approx 8.33 \, \text{Hz}$) and a sampling frequency $f_s$ of $51.2 \, \text{kHz}$. It can be noticed that the spectrogram exhibits distinct periodical vertical lines. The corresponding \ac{AMS} is shown in the lower subfigure of Figure~\ref{fig:ams_damaged} where the subbands of the first \ac{STFT} are given along the y-axis, and the modulation frequencies are illustrated along the x-axis. It can be observed that for center frequencies above $20$~kHz strong modulations are present. This is, however, not the case for the AMS of the vibration signals of a healthy bearing operated under identical operational conditions (cf. Figure~\ref{fig:ams_healthy}). The aforementioned distinct vertical lines of the spectrogram are especially noticeable in the high frequencies above $15$~kHz (highlighted by dotted red lines) and are well captured by the energy in the AMS. One might argue that inspecting the spectrograms in Figure~\ref{fig:ams_damaged} and Figure~\ref{fig:ams_healthy} is sufficient for detecting the fault. However, the \ac{AMS} provides a larger energy difference in the considered frequency range. In order to quantify these differences with a single scalar value, we propose to sum up the amplitudes for center frequencies above $20 \, \text{kHz}$ and modulation frequencies below $80 \, \text{Hz}$.
\begin{table}[h]
    \centering
    \caption{Parameters for the computation of the two STFTs. The number of samples is given for the sampling frequency of $f_s = 51.2 \,\text{kHz}$. }
    \begin{tabular}{c c c}
     \textbf{Parameter} & \textbf{1. STFT}  & \textbf{2. STFT}  \\
      \hline
      Window length &  25 ms (1280 samples)  &  512 ms  (128 samples)  \\ \hline
      Step size & 4 ms ( 205 samples)  &  256 ms (64 samples)  \\
      \hline 
      N-FFT  &512  & 256 \\
      \hline 
    \end{tabular}
\end{table}
   \label{Tab:param_nfft}

\subsection{One-Class SVM}
\label{sec:svm}
\noindent
The One-Class \ac{SVM} represents a flexible and effective method for the detection of outliers. It was originally introduced by Sch\"olkopf et al.  \cite{schoelkopf2001} in the context of novelty detection and aims at estimating a boundary function which returns positive values for data points belonging to a given class. For data points of other classes and outliers, the function returns negative values. 

In the following, we will briefly review the fundamental idea of One-Class \acp{SVM}. For this, we consider a training data set $\{ \mathbf{x}_{n}\}_{n=1, \ldots, N_{}}$ belonging to one class, where $N$ is the overall number of training samples. 
The boundary function is estimated in a so-called kernel-induced feature space where the training data is assumed to be linearly separable from the origin. An according decision boundary is then given by 
\begin{align}
   f(\mathbf{x}) = \mathbf{w}^\mathrm{T} \phi(\mathbf{x}) + b, 
\end{align}
where $b$ and $\mathbf{\mathbf{w}}$ denote the distance from the origin and the normal vector, respectively. $\phi(\cdot)$ is the mapping function to the kernel-induced feature space. Similar to conventional \acp{SVM}, $\phi(\cdot)$ does not have to be known explicitly what is generally referred to as kernel-trick \cite{bishop2006}. The parameters of $f(\cdot)$ must be optimized such that the training data is separated from the origin with maximal distance and $f(\mathbf{x}_n) \geq 0$ holds for all $n$. 
Considering that the Euclidean distance from any arbitrary point $\mathbf{x}$ to $f(\cdot)$ is given by $f(\mathbf{x})/\| \mathbf{w} \|_2$ and ensuring that $\phi(0)=0$ holds, a boundary with maximal distance from the origin can be obtained by solving the following optimization problem:
\begin{align}
    &\underset{\mathbf{w},~b}{\text{minimize}} \quad \| \mathbf{w} \|_2 -b     \label{eq:optproblem} \\
    &\text{subject to} \quad  \mathbf{\mathbf{w}}^\mathrm{T} \phi(\mathbf{x}_n) + b \geq 0, \forall n. \nonumber
\end{align}
Equation~\ref{eq:optproblem} can be solved using methods from constrained optimization, e.g., solving the Lagrangian \cite{bishop2006}. Similar to conventional SVMs, the hyperplane $f(\cdot)$ will be  defined by only some data points from the training set, the so-called support vectors. Furthermore, it is shown in \cite{schoelkopf2001} that the obtained decision boundary is an estimator for the support of a density that generated the training data points. Note that Equation~\ref{eq:optproblem} can be extended by slack variables to trade-off the width of the margin and the number of misclassified data points, cf. \cite{schoelkopf2001}. 

As mentioned above, the implicit mapping of  $\mathbf{x}$ by  $\phi (\cdot)$  does not have to be known explicitly as we are only interested in the inner products of the mapped inputs, which can be computed using a kernel. Usually, One-Class \acp{SVM} operate with a Gaussian kernel $e^{-\gamma \| \mathbf{x}_i - \mathbf{x}_j \|}$ for data points $\mathbf{x}_i$ and $\mathbf{x}_j$. The Gaussian kernel has only one free parameter $\gamma$ which is referred to as the \textit{width} of the kernel. Note that the Gaussian kernel ensures the existence of a hyperplane defined by the One-Class SVM optimization problem \cite{schoelkopf2001}.

\section{Evaluation}
\label{sec:evaluation}
\noindent
In the following, the experimental setup (cf. Section~\ref{sec:experimental_setup}), the data generation and training process as well as the considered evaluation metrics (cf. Section~\ref{sec:data}) are described, before the obtained results are thoroughly discussed in Section~\ref{sec:results}.

\subsection{Experimental Setup}
\label{sec:experimental_setup}
\noindent
In the experimental study, a state-of-the-art commuter rail induction engine with a  nominal capacity of $405 \, \text{kW}$, a nominal load of $1800 \,\text{Nm}$ and a nominal rotation speed of $2150 \, \text{rpm}$ is studied. In total, $3$ different bearings are investigated, including a healthy bearing and two bearings with different types of damages. One of the bearings exhibits an artificially introduced damage in the outer race with a fault diameter of approximately $3.5 \, \text{mm}$ and a  depth of approximately $1.5 \, \text{mm}$. The width of the outer race is  $22 \, \text{mm}$. This can be considered  as a localized fault in a rather early stage. The second bearing returned from actually usage in the field in a railway vehicle after a run-time of approximately $1.5$~million \text{km}, which exhibits distributed faults at multiple bearing components including the outer race, the rolling elements, the cage and the rolling elements. This bearing represents a naturally developed bearing damage in an advanced development stage. In the following these two damage patterns are referred to as artificially and naturally developed damage, respectively. The investigated bearings (type 6016) are installed at the drive-end position of the induction engine.

The experimental setup, which is shown in Figure~\ref{fig:setup}, further includes a second induction engine of identical type that is used as a load machine in a back-to-back setup. In total, two acceleration sensors (Sensor A and Sensor B, $f_s = 51,2 \, \text{kHz}$) are mounted on the housing of the induction engine at the drive-end and the non-drive-end position, respectively. An additional sensor is used to track the rotational speed. Further, a railway traction converter is used in the experiments, which strongly influences the vibration pattern of the induction engine. The applied torque lies in the interval $-2900 \, \text{Nm}$ and $2900 \, \text{Nm}$ with four different torque levels (0~\%, 33~\%, 66~\%, 100~\%) for accelerating and breaking. The applied rotational speed follows a steplike characteristic with 8 distinct rotational speed levels: $500$~rpm, $750$~rpm, $1000$~rpm, $1500$~rpm, $2000$~rpm, $2500$~rpm, $3000$~rpm and $3500$~rpm. The rotational speed curve is shown in Figure~\ref{fig:speed_curve}. It can be observed that the rotational speed was kept constant for a period of approximately $60 \, \text{s}$ for every rotational speed. Overall, the examined scenario is highly challenging and close to the operating conditions in the field as the rotational speed and the torque are strongly varying, which is not the case  for publicly available datasets as, e.g., the Case Western Reserve University (CWRU) dataset \cite{neupane2020}.
\begin{figure}
\centering
\begin{minipage}[b]{.5\textwidth}
  \centering
  \input{figures/measurement_setup.tex}
  \caption{Experimental setup.}
  \label{fig:setup}
\end{minipage}%
\begin{minipage}[b]{.5\textwidth}
  \centering
  \setlength{\fwidth}{6cm}
  \setlength{\fheight}{3.5cm}
  \input{figures/a1_speedcurve}
  \caption{Rotational speed curve.}
  \label{fig:speed_curve}
\end{minipage}
\end{figure}

\subsection{Data Generation, Training, and Evaluation Metrics}
\label{sec:data}
\noindent 
For generating the training set and the test set, the vibration signals were divided into non-overlapping frames with a length of $2 \, \text{s}$ each. After segmentation, frames containing transitions of the rotational frequency were dismissed. For both sensors, this resulted in 1558 frames for the healthy measurement, 2512 frames for the artificially damaged bearing and 2728 frames for the naturally developed damage. For training the One-Class \ac{SVM}, 500 samples obtained from the healthy bearing were randomly chosen for the training set. Another small dataset containing 50 additional samples from the measurements of the healthy bearing was used as an evaluation dataset for optimizing the parameters of the One-Class \ac{SVM} using a grid-search procedure.

In the following, the evaluation metrics that are considered for rating the classification performance are introduced. We consider a binary classification problem where we distinguish between the class P (positive, healthy bearing) and the class N (negative, damaged bearing). The performance of a classifier can be assessed in an illustrative manner with a so-called confusion matrix (CM) (cf. Figure \ref{fig:cm}). In a CM, the predicted class labels (y-axis) are compared to the true class labels (x-axis). Hence, the diagonal entries, i.e., True Positive (TP) and True Negative (TN), show the number of correctly classified test instances, whereas the off-diagonal entries, i.e., False Negative (FN) and False Positive (FP), represent the false predictions. Applied to the classification of bearing faults, the FN value corresponds to the number of healthy bearings that were falsely classified as damaged and vice-versa for FP. By construction our test dataset is highly imbalanced as the number of negative test instances is much greater than the number of positive test instances. Consequently, metrics such as the accuracy or the F1-score are not suited and we consider the True Positive Rate (TPR) and the True Negative Rate (TNR), which specify how well the positive class and the negative class were predicted, respectively, instead. The arithmetic mean of the TPR and the TNR leads to the Balanced Accuracy (BA). The formulaes for computing TPR, TNR and BA are summarized in Table~\ref{tab:metrics}. Further, the False Negative Rate (FNR) and the False Positive Rate (FPR) denote the fraction of the test instances that were misclassified as P or N and are given as $1-\text{FNR}$ and $1-\text{FPR}$. 
\begin{figure}
\begin{minipage}[t]{.5\textwidth}
    \begin{center}
    \input{figures/confusion_matrix}
    \end{center}
    \captionsetup{type=figure,format=hang}
    \caption{\sloppy{Confusion matrix for a binary $\qquad$ classifi\-cation problem.}}
    \label{fig:cm}
\end{minipage}
\begin{minipage}[b]{.5\textwidth}
    \centering
     \captionsetup{type=table}
    \caption{Evaluation metrics.}
    \begin{tabular}{c|c}
    \textbf{Name}     & \textbf{Formula}  \\
    \hline 
    TPR (True Positive Rate)  &  $\frac{\text{TP}}{\text{TP}+\text{FN}}$\\
    \hline
    TNR (True Negative Rate) & $\frac{\text{TN}}{\text{TN}+\text{FP}} $\\
    \hline
    BA (Balanced Accuracy) & $\frac{\text{TPR}+\text{TNR}}{2} $\\
    \hline
    \end{tabular}
     \label{tab:metrics}
\end{minipage}
\end{figure}
\subsection{Results}
\label{sec:results}
\begin{figure}[bth]
    \centering
    \setlength{\fwidth}{0.75\textwidth}
    \setlength{\fheight}{3.5cm}
    \input{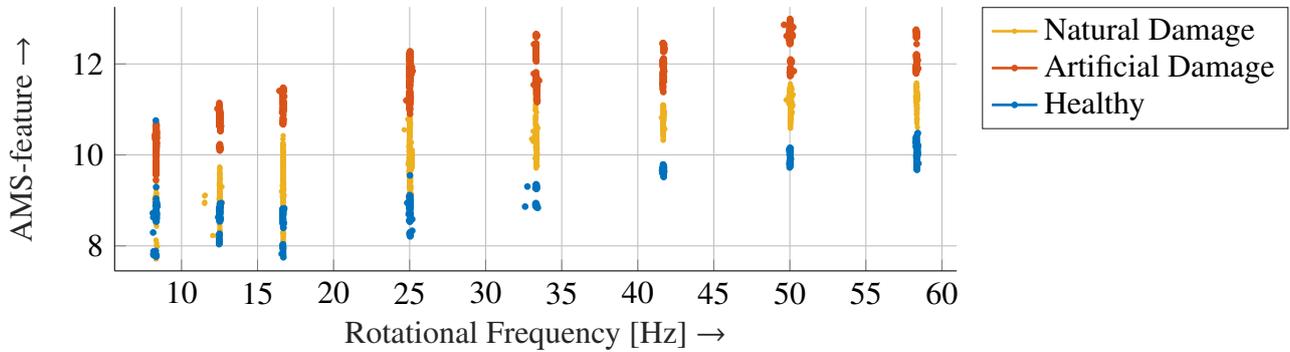}
    \caption{AMS feature in dependency of the rotational frequency for healthy (blue),
    artificially (red) and naturally damaged (yellow) bearing.}
    \label{fig:ams_feat_comparison}
\end{figure}

\noindent
For evaluation, the One-Class \ac{SVM} was trained using the five different feature sets that were introduced in Sections~\ref{sec:features}-\ref{sec:mfcc}: the time-domain features (\textit{TD}), the spectral descriptors (\textit{SD}), the amplitudes at the characteristic fault frequencies in the envelope spectrum (\textit{ENV-AMP}), the scalar feature derived from the Amplitude Modulation Spectrogram (\textit{AMS}) and the first 13 Mel Frequency Spectral Coefficients (\textit{MFCC (13)}). Further, to investigate whether the addition of the rotational frequency $f_r$ is beneficial, all feature sets were again evaluated after appending $f_r$ to the feature vectors. Training and testing were repeated 10 times using different realizations of the training and test set. Averaged results for TPR, TNR and BA are summarized in Table~\ref{Tab:Res1} for Sensor A. Overall, it can be stated that the feature set \textit{MFCC (13)} leads to the best classification results with a BA of 98.79~\% and 98.91~\% after appending the rotational frequency to the feature set. BAs over 90~\% are only achieved for the \textit{ENV AMP +} $f_r$ and \textit{AMS+}${f_r}$ with 91.58~\% and 94.81~\%, respectively, whereas the \textit{TD} and \textit{SD} feature performs worst. Surprisingly, a large performance drop can be witnessed when $f_r$ is added to \textit{SD} (from 88.33~\% to 82.81~\% in terms of BA). The reasons for this remains to investigated. However, for the other feature sets, the addition of $f_r$ to the feature vector improves the classification results always at least slightly. Yet, the TPR does not always improve, which calls for further investigation.       The most noticeable difference can be observed for \textit{AMS} where the BA improves from 85.13~\% to
94.81~\%. This steep rise in performance can be explained by inspecting Figure~\ref{fig:ams_feat_comparison} where the
AMS feature is plotted for different $f_r$ and healthy (blue), artificially damaged (red) and naturally damaged (yellow) bearings. A clear margin between the artificially damaged and the healthy bearing can be observed. The \ac{AMS} feature values for the naturally damaged bearing populate the region between the two other classes. However, for $f_r > 30 \, \text{Hz}$ also a clear distinction can be made between the naturally damaged and the healthy bearing. The observation that the \ac{AMS} feature is well-suited for the classification of the artificially damaged bearing is
further validated by a BA of 99.65~\%, when only test instances from the artificially damaged bearing are considered in the test set. However, the classification performance increases also significantly for all other feature sets in this scenario. 

\begin{figure}[hbt!]
    \setlength{\fwidth}{0.75\textwidth}
    \setlength{\fheight}{3.5cm}
    \input{figures/mfcc_sensorB_only_healthy}
    \caption{Accuracy, FPR, FNR and BA in dependency of the number of MFCCs for
Sensor B}
    \label{fig:mfcc_sensor_b}
\end{figure}
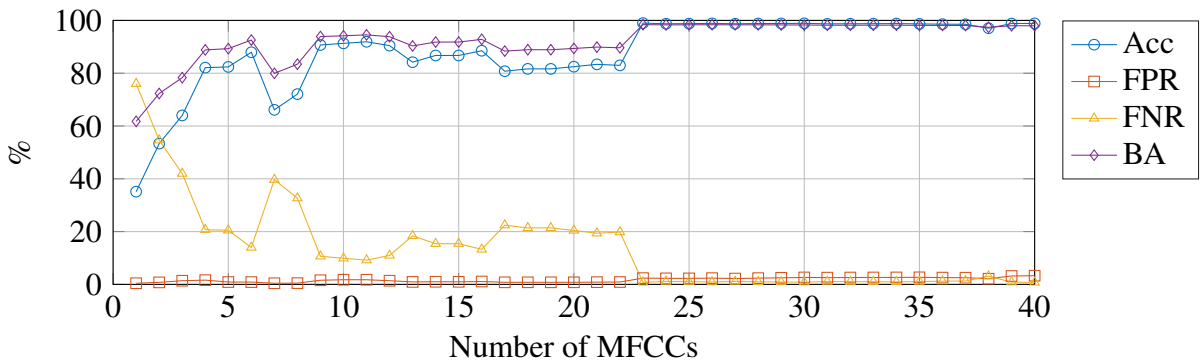

Inspecting the results for Sensor B (cf. Table~\ref{Tab:Res2}), which is located at the opposite end of the induction engine, it becomes evident that the larger distance between bearing and sensor heavily impacts the classification results. From  Table~\ref{Tab:Res2}, a significant drop in performance by approximately 10 percentage points for \textit{TD} and \textit{SD} and 7 percentage points for \textit{ENV AMP} and \textit{AMS} in terms of BA can be observed. The TNR drops significantly, i.e., many bearing faults remain undetected and are misclassified as healthy. Solely the \acp{MFCC} and to some extent the \textit{AMS} +  ${f_r}$ feature sets yield high TNRs. 
As the \acp{MFCC} have proven to be the most effective features, it was further investigated how the performance changes with a varying number of \acp{MFCC}. Accuracy, FPR, FNR and BA in dependence of the number of used \acp{MFCC} for Sensor B are shown in Figure~\ref{fig:mfcc_sensor_b}. In Figure~\ref{fig:mfcc_sensor_b} a significant performance gain can be observed for Sensor B when the number of included \acp{MFCC} rises above 23 and the FNR, the number of healthy bearings that are misclassified, drops noticeably. Yet, for Sensor A, very good classification results can already be obtained after using only the first five \acp{MFCC}. A similar behavior can also be observed for Sensor B, when a few instances of damaged bearings are also included in the evaluation dataset that is used for optimizing the \ac{SVM} parameters. The use of such an evaluation dataset did not, however, improve the classification performance for the other feature sets to the same extent.

\begin{table}
\begin{minipage}[b]{.5\textwidth}
\caption{Classification results for Sensor A.}
\label{Tab:Res1}
\begin{center}
\begin{tabular}{l  c c c} 
    \hline
    \textbf{Feature Set} &  \textbf{TPR} & \textbf{TNR} & \textbf{BA} \\ [0.5ex] 
    \hline
    TD & 97.81 &   80.51 &  89.16  \\
    \hline 
    TD + $f_{\text{r}}$ &   97.22 &  82.58 &  89.90 \\
    \hline
    SD & 98.53 &   78.13 &  88.33 \\
    \hline
    SD + $f_{\text{r}}$ & 99.34 &   66.29 &   82.81\\
    \hline
    ENV AMP &   85.55  & 95.32 &  90.44 \\
    \hline 
    ENV AMP + $ f_{\text{r}}$ & 86.73 &   96.44 &   91.58 \\ 
    \hline
    AMS &   97.35  & 72.92 &  85.13 \\
    \hline
    AMS + $f_{\text{r}}$ & 99.33 &  90.29 &   94.81\\ 
    \hline
    MFCC  & 98.23 &   99.34 &   98.79\\
    \hline 
    MFCC  + $f_{\text{r}}$ & 98.20 &   99.62 &   98.91\\
    \hline
 \end{tabular}
 \end{center}
\end{minipage}
\begin{minipage}[b]{.5\textwidth}
\caption{Classification results for Sensor B.}
\label{Tab:Res2}
\begin{center}
\begin{tabular}{l c c c} 
    \hline
    \textbf{Feature Set} &  \textbf{TPR} & \textbf{TNR} & \textbf{BA} \\ [0.5ex] 
    \hline
    TD &  96.67 &   60.91 &   78.79 \\
    \hline 
    TD + $f_{\text{r}}$ &   97.57 &   62.11 &  79.84  \\
    \hline
    SD & 97.66 &   65.37 &   81.51 \\
    \hline
    SD + $f_{\text{r}}$ & 99.34 &   66.29 &   82.81\\
    \hline
    ENV AMP &  96.65 &   71.99 &   84.32 \\
    \hline 
    ENV AMP + $ f_{\text{r}}$ & 97.96  & 71.24 &  84.60  \\ 
    \hline
    AMS &    94.19  & 50.37  & 72.28\\
    \hline
    AMS + $f_{\text{r}}$ & 91.15 &   82.95 &   87.05\\  
    \hline
    MFCC  & 98.30 &   94.39 &  96.35\\
    \hline 
    MFCC  + $f_{\text{r}}$ & 98.25 &   95.32 &   96.79\\
    \hline
 \end{tabular}
 \end{center}
 \end{minipage}
 \end{table}


\section{Summary}
\label{sec:summary}
\noindent
In this article, well-established features for the detection of bearing faults together with state-of-the-art features from the audio signal processing domain were investigated. In our approach, bearing faults were classified following an anomaly detection approach in which a One-Class \ac{SVM} was trained solely with data obtained from healthy bearings. Extensive experiments, conducted on a state-of-the-art commuter railway induction engine in a challenging scenario, proved that the \acp{MFCC} and the \ac{AMS} feature can  be highly effective features for the detection of bearing faults.


\bibliographystyle{unsrt}
\bibliography{references} 
\end{document}

%% file: figures/ams.tex
\usetikzlibrary{calc}

\begin{tikzpicture}[scale=0.8,every node/.style={scale=0.9}]

\node[,color=black, align = center, text= black, minimum width = 0.5cm] at (0,0) (x) {$x[k]$};

\node[draw,color=black, align = center, text= black, minimum width = 3.25cm,rotate=90] at (2,0) (stft) {STFT};

\draw[->,thick] (x) -- (stft);
\node[draw,color=black, align = center, text= black, minimum width = 1cm] at (3.5,1.25) (absu) {$\left| \cdot \right|^2$};

\node[draw,color=black, align = center, text= black, minimum width = 1cm] at (3.5,-1.25) (absd) {$\left| \cdot \right|^2$};
\draw[->,thick] ($(stft) + (0.3,1.25)$) -- (absu);
\draw[->,thick] ($(stft) + (0.3,-1.25)$) -- (absd);
\node[color=black, align = center, text= black, minimum width = 1cm,rotate=90] at (3.5,0) (dots1) {$\cdots$};

\node[draw,color=black, align = center, text= black, minimum width = 1cm,minimum height=0.75cm] at (6.0,1.25) (stft2u) {STFT};
\node[draw,color=black, align = center, text= black, minimum width = 1cm,minimum height=0.75cm] at (6.0,-1.25) (stft2d) {STFT};
\node[color=black, align = center, text= black, minimum width = 1cm,rotate=90] at (5.5,0) (dots2) {$\cdots$};
\draw[->,thick] (absu) -- (stft2u);
\draw[->,thick] (absd) -- (stft2d);

\node[draw,color=black, align = center, text= black, minimum width = 1cm] at (8.0,1.25) (abs2u) {$\left| \cdot \right|$};
\node[draw,color=black, align = center, text= black, minimum width = 1cm] at (8.0,-1.25) (abs2d) {$\left| \cdot \right|$};
\node[color=black, align = center, text= black, minimum width = 1cm,rotate=90] at (8,0) (dots3) {$\cdots$};
\draw[->,thick] ($(stft2u.east) + (0,0.25)$) --( $ (abs2u.west) + (0,0.25)$);
\draw[->,thick] ($(stft2u.east) + (0,-0.25)$) --( $(abs2u.west)+ (0,-0.25)$ );
\node[color=black, align = center, text= black, minimum width = 1cm,rotate=90] at (7,1.25) (dots4u)  {\scriptsize $\cdots$};
\node[color=black, align = center, text= black, minimum width = 1cm,rotate=90] at (7,-1.25) (dots4d)  {\scriptsize $\cdots$};

\draw[->,thick] ($(stft2d.east) + (0,0.25)$) --( $ (abs2d.west) + (0,0.25)$);
\draw[->,thick] ($(stft2d.east) + (0,-0.25)$) --( $(abs2d.west)+ (0,-0.25)$ );

\node[draw,color=black, align = center, text= black, minimum width = 1cm, minimum height=0.75cm] at (9.5,1.25) (dBu) {log};
\node[draw,color=black, align = center, text= black, minimum width = 1cm,minimum height=0.75cm] at (9.5,-1.25) (dBd) {log};

\draw[->,thick] ($(abs2u.east) + (0,0.25)$) --( $ (dBu.west) + (0,0.25)$);
\draw[->,thick] ($(abs2u.east) + (0,-0.25)$) --( $(dBu.west)+ (0,-0.25)$ );
\draw[->,thick] ($(abs2d.east) + (0,0.25)$) --( $ (dBd.west) + (0,0.25)$);
\draw[->,thick] ($(abs2d.east) + (0,-0.25)$) --( $(dBd.west)+ (0,-0.25)$ );
\node[draw,color=black, align = center, text= black, minimum width = 3.25cm,rotate=90] at (11.5,0) (ams) {AMS};



\draw[->,thick] ($(dBd.east) + (0,0.25)$) --( $ (ams.north) + (0,-1)$);
\draw[->,thick] ($(dBd.east) + (0,-0.25)$) --( $(ams.north)+ (0,-1.5)$ );

\draw[->,thick] ($(dBu.east) + (0,0.25)$) --( $ (ams.north) + (0,1.5)$);
\draw[->,thick] ($(dBu.east) + (0,-0.25)$) --( $(ams.north)+ (0,1.0)$ );

\node[color=black, align = center, text= black, minimum width = 1cm,rotate=90] at (8.75,1.25) (dots4u)  {\scriptsize $\cdots$};
\node[color=black, align = center, text= black, minimum width = 1cm,rotate=90] at (8.75,-1.25) (dots4d)  {\scriptsize $\cdots$};
\node[color=black, align = center, text= black, minimum width = 1cm,rotate=90] at (9.5,0) () {$\cdots$};

\node[color=black, align = center, text= black, minimum width = 1cm,rotate=90] at (10.5,1.25) (dots4u)  {\scriptsize $\cdots$};
\node[color=black, align = center, text= black, minimum width = 1cm,rotate=90] at (10.5,-1.25) (dots4d)  {\scriptsize $\cdots$};
\end{tikzpicture}

%% file: figures/ams_damaged.tex
%
%
\begin{tikzpicture}
\tikzstyle{every node}=[font=\scriptsize]
\begin{axis}[%
width=0.842\fwidth,
height=0.406\fheight,
at={(0\fwidth,0.594\fheight)},
scale only axis,
point meta min=-35.6596843486583,
point meta max=63.0850781322591,
axis on top,
xmin=0.01,
xmax=1.98879078694818,
xlabel style={font=\color{white!15!black},font=\scriptsize},
xlabel={Time [s] $\rightarrow$},
ymin=-50.87890625,
ymax=26100.87890625,
ylabel style={font=\color{white!15!black},font=\scriptsize},
ylabel={Frequency [Hz] $\rightarrow$},
axis background/.style={fill=white},
legend style={legend cell align=left, align=left, draw=white!15!black},
colormap={mymap}{[1pt] rgb(0pt)=(0.2422,0.1504,0.6603); rgb(1pt)=(0.2444,0.1534,0.6728); rgb(2pt)=(0.2464,0.1569,0.6847); rgb(3pt)=(0.2484,0.1607,0.6961); rgb(4pt)=(0.2503,0.1648,0.7071); rgb(5pt)=(0.2522,0.1689,0.7179); rgb(6pt)=(0.254,0.1732,0.7286); rgb(7pt)=(0.2558,0.1773,0.7393); rgb(8pt)=(0.2576,0.1814,0.7501); rgb(9pt)=(0.2594,0.1854,0.761); rgb(11pt)=(0.2628,0.1932,0.7828); rgb(12pt)=(0.2645,0.1972,0.7937); rgb(13pt)=(0.2661,0.2011,0.8043); rgb(14pt)=(0.2676,0.2052,0.8148); rgb(15pt)=(0.2691,0.2094,0.8249); rgb(16pt)=(0.2704,0.2138,0.8346); rgb(17pt)=(0.2717,0.2184,0.8439); rgb(18pt)=(0.2729,0.2231,0.8528); rgb(19pt)=(0.274,0.228,0.8612); rgb(20pt)=(0.2749,0.233,0.8692); rgb(21pt)=(0.2758,0.2382,0.8767); rgb(22pt)=(0.2766,0.2435,0.884); rgb(23pt)=(0.2774,0.2489,0.8908); rgb(24pt)=(0.2781,0.2543,0.8973); rgb(25pt)=(0.2788,0.2598,0.9035); rgb(26pt)=(0.2794,0.2653,0.9094); rgb(27pt)=(0.2798,0.2708,0.915); rgb(28pt)=(0.2802,0.2764,0.9204); rgb(29pt)=(0.2806,0.2819,0.9255); rgb(30pt)=(0.2809,0.2875,0.9305); rgb(31pt)=(0.2811,0.293,0.9352); rgb(32pt)=(0.2813,0.2985,0.9397); rgb(33pt)=(0.2814,0.304,0.9441); rgb(34pt)=(0.2814,0.3095,0.9483); rgb(35pt)=(0.2813,0.315,0.9524); rgb(36pt)=(0.2811,0.3204,0.9563); rgb(37pt)=(0.2809,0.3259,0.96); rgb(38pt)=(0.2807,0.3313,0.9636); rgb(39pt)=(0.2803,0.3367,0.967); rgb(40pt)=(0.2798,0.3421,0.9702); rgb(41pt)=(0.2791,0.3475,0.9733); rgb(42pt)=(0.2784,0.3529,0.9763); rgb(43pt)=(0.2776,0.3583,0.9791); rgb(44pt)=(0.2766,0.3638,0.9817); rgb(45pt)=(0.2754,0.3693,0.984); rgb(46pt)=(0.2741,0.3748,0.9862); rgb(47pt)=(0.2726,0.3804,0.9881); rgb(48pt)=(0.271,0.386,0.9898); rgb(49pt)=(0.2691,0.3916,0.9912); rgb(50pt)=(0.267,0.3973,0.9924); rgb(51pt)=(0.2647,0.403,0.9935); rgb(52pt)=(0.2621,0.4088,0.9946); rgb(53pt)=(0.2591,0.4145,0.9955); rgb(54pt)=(0.2556,0.4203,0.9965); rgb(55pt)=(0.2517,0.4261,0.9974); rgb(56pt)=(0.2473,0.4319,0.9983); rgb(57pt)=(0.2424,0.4378,0.9991); rgb(58pt)=(0.2369,0.4437,0.9996); rgb(59pt)=(0.2311,0.4497,0.9995); rgb(60pt)=(0.225,0.4559,0.9985); rgb(61pt)=(0.2189,0.462,0.9968); rgb(62pt)=(0.2128,0.4682,0.9948); rgb(63pt)=(0.2066,0.4743,0.9926); rgb(64pt)=(0.2006,0.4803,0.9906); rgb(65pt)=(0.195,0.4861,0.9887); rgb(66pt)=(0.1903,0.4919,0.9867); rgb(67pt)=(0.1869,0.4975,0.9844); rgb(68pt)=(0.1847,0.503,0.9819); rgb(69pt)=(0.1831,0.5084,0.9793); rgb(70pt)=(0.1818,0.5138,0.9766); rgb(71pt)=(0.1806,0.5191,0.9738); rgb(72pt)=(0.1795,0.5244,0.9709); rgb(73pt)=(0.1785,0.5296,0.9677); rgb(74pt)=(0.1778,0.5349,0.9641); rgb(75pt)=(0.1773,0.5401,0.9602); rgb(76pt)=(0.1768,0.5452,0.956); rgb(77pt)=(0.1764,0.5504,0.9516); rgb(78pt)=(0.1755,0.5554,0.9473); rgb(79pt)=(0.174,0.5605,0.9432); rgb(80pt)=(0.1716,0.5655,0.9393); rgb(81pt)=(0.1686,0.5705,0.9357); rgb(82pt)=(0.1649,0.5755,0.9323); rgb(83pt)=(0.161,0.5805,0.9289); rgb(84pt)=(0.1573,0.5854,0.9254); rgb(85pt)=(0.154,0.5902,0.9218); rgb(86pt)=(0.1513,0.595,0.9182); rgb(87pt)=(0.1492,0.5997,0.9147); rgb(88pt)=(0.1475,0.6043,0.9113); rgb(89pt)=(0.1461,0.6089,0.908); rgb(90pt)=(0.1446,0.6135,0.905); rgb(91pt)=(0.1429,0.618,0.9022); rgb(92pt)=(0.1408,0.6226,0.8998); rgb(93pt)=(0.1383,0.6272,0.8975); rgb(94pt)=(0.1354,0.6317,0.8953); rgb(95pt)=(0.1321,0.6363,0.8932); rgb(96pt)=(0.1288,0.6408,0.891); rgb(97pt)=(0.1253,0.6453,0.8887); rgb(98pt)=(0.1219,0.6497,0.8862); rgb(99pt)=(0.1185,0.6541,0.8834); rgb(100pt)=(0.1152,0.6584,0.8804); rgb(101pt)=(0.1119,0.6627,0.877); rgb(102pt)=(0.1085,0.6669,0.8734); rgb(103pt)=(0.1048,0.671,0.8695); rgb(104pt)=(0.1009,0.675,0.8653); rgb(105pt)=(0.0964,0.6789,0.8609); rgb(106pt)=(0.0914,0.6828,0.8562); rgb(107pt)=(0.0855,0.6865,0.8513); rgb(108pt)=(0.0789,0.6902,0.8462); rgb(109pt)=(0.0713,0.6938,0.8409); rgb(110pt)=(0.0628,0.6972,0.8355); rgb(111pt)=(0.0535,0.7006,0.8299); rgb(112pt)=(0.0433,0.7039,0.8242); rgb(113pt)=(0.0328,0.7071,0.8183); rgb(114pt)=(0.0234,0.7103,0.8124); rgb(115pt)=(0.0155,0.7133,0.8064); rgb(116pt)=(0.0091,0.7163,0.8003); rgb(117pt)=(0.0046,0.7192,0.7941); rgb(118pt)=(0.0019,0.722,0.7878); rgb(119pt)=(0.0009,0.7248,0.7815); rgb(120pt)=(0.0018,0.7275,0.7752); rgb(121pt)=(0.0046,0.7301,0.7688); rgb(122pt)=(0.0094,0.7327,0.7623); rgb(123pt)=(0.0162,0.7352,0.7558); rgb(124pt)=(0.0253,0.7376,0.7492); rgb(125pt)=(0.0369,0.74,0.7426); rgb(126pt)=(0.0504,0.7423,0.7359); rgb(127pt)=(0.0638,0.7446,0.7292); rgb(128pt)=(0.077,0.7468,0.7224); rgb(129pt)=(0.0899,0.7489,0.7156); rgb(130pt)=(0.1023,0.751,0.7088); rgb(131pt)=(0.1141,0.7531,0.7019); rgb(132pt)=(0.1252,0.7552,0.695); rgb(133pt)=(0.1354,0.7572,0.6881); rgb(134pt)=(0.1448,0.7593,0.6812); rgb(135pt)=(0.1532,0.7614,0.6741); rgb(136pt)=(0.1609,0.7635,0.6671); rgb(137pt)=(0.1678,0.7656,0.6599); rgb(138pt)=(0.1741,0.7678,0.6527); rgb(139pt)=(0.1799,0.7699,0.6454); rgb(140pt)=(0.1853,0.7721,0.6379); rgb(141pt)=(0.1905,0.7743,0.6303); rgb(142pt)=(0.1954,0.7765,0.6225); rgb(143pt)=(0.2003,0.7787,0.6146); rgb(144pt)=(0.2061,0.7808,0.6065); rgb(145pt)=(0.2118,0.7828,0.5983); rgb(146pt)=(0.2178,0.7849,0.5899); rgb(147pt)=(0.2244,0.7869,0.5813); rgb(148pt)=(0.2318,0.7887,0.5725); rgb(149pt)=(0.2401,0.7905,0.5636); rgb(150pt)=(0.2491,0.7922,0.5546); rgb(151pt)=(0.2589,0.7937,0.5454); rgb(152pt)=(0.2695,0.7951,0.536); rgb(153pt)=(0.2809,0.7964,0.5266); rgb(154pt)=(0.2929,0.7975,0.517); rgb(155pt)=(0.3052,0.7985,0.5074); rgb(156pt)=(0.3176,0.7994,0.4975); rgb(157pt)=(0.3301,0.8002,0.4876); rgb(158pt)=(0.3424,0.8009,0.4774); rgb(159pt)=(0.3548,0.8016,0.4669); rgb(160pt)=(0.3671,0.8021,0.4563); rgb(161pt)=(0.3795,0.8026,0.4454); rgb(162pt)=(0.3921,0.8029,0.4344); rgb(163pt)=(0.405,0.8031,0.4233); rgb(164pt)=(0.4184,0.803,0.4122); rgb(165pt)=(0.4322,0.8028,0.4013); rgb(166pt)=(0.4463,0.8024,0.3904); rgb(167pt)=(0.4608,0.8018,0.3797); rgb(168pt)=(0.4753,0.8011,0.3691); rgb(169pt)=(0.4899,0.8002,0.3586); rgb(170pt)=(0.5044,0.7993,0.348); rgb(171pt)=(0.5187,0.7982,0.3374); rgb(172pt)=(0.5329,0.797,0.3267); rgb(173pt)=(0.547,0.7957,0.3159); rgb(175pt)=(0.5748,0.7929,0.2941); rgb(176pt)=(0.5886,0.7913,0.2833); rgb(177pt)=(0.6024,0.7896,0.2726); rgb(178pt)=(0.6161,0.7878,0.2622); rgb(179pt)=(0.6297,0.7859,0.2521); rgb(180pt)=(0.6433,0.7839,0.2423); rgb(181pt)=(0.6567,0.7818,0.2329); rgb(182pt)=(0.6701,0.7796,0.2239); rgb(183pt)=(0.6833,0.7773,0.2155); rgb(184pt)=(0.6963,0.775,0.2075); rgb(185pt)=(0.7091,0.7727,0.1998); rgb(186pt)=(0.7218,0.7703,0.1924); rgb(187pt)=(0.7344,0.7679,0.1852); rgb(188pt)=(0.7468,0.7654,0.1782); rgb(189pt)=(0.759,0.7629,0.1717); rgb(190pt)=(0.771,0.7604,0.1658); rgb(191pt)=(0.7829,0.7579,0.1608); rgb(192pt)=(0.7945,0.7554,0.157); rgb(193pt)=(0.806,0.7529,0.1546); rgb(194pt)=(0.8172,0.7505,0.1535); rgb(195pt)=(0.8281,0.7481,0.1536); rgb(196pt)=(0.8389,0.7457,0.1546); rgb(197pt)=(0.8495,0.7435,0.1564); rgb(198pt)=(0.86,0.7413,0.1587); rgb(199pt)=(0.8703,0.7392,0.1615); rgb(200pt)=(0.8804,0.7372,0.165); rgb(201pt)=(0.8903,0.7353,0.1695); rgb(202pt)=(0.9,0.7336,0.1749); rgb(203pt)=(0.9093,0.7321,0.1815); rgb(204pt)=(0.9184,0.7308,0.189); rgb(205pt)=(0.9272,0.7298,0.1973); rgb(206pt)=(0.9357,0.729,0.2061); rgb(207pt)=(0.944,0.7285,0.2151); rgb(208pt)=(0.9523,0.7284,0.2237); rgb(209pt)=(0.9606,0.7285,0.2312); rgb(210pt)=(0.9689,0.7292,0.2373); rgb(211pt)=(0.977,0.7304,0.2418); rgb(212pt)=(0.9842,0.733,0.2446); rgb(213pt)=(0.99,0.7365,0.2429); rgb(214pt)=(0.9946,0.7407,0.2394); rgb(215pt)=(0.9966,0.7458,0.2351); rgb(216pt)=(0.9971,0.7513,0.2309); rgb(217pt)=(0.9972,0.7569,0.2267); rgb(218pt)=(0.9971,0.7626,0.2224); rgb(219pt)=(0.9969,0.7683,0.2181); rgb(220pt)=(0.9966,0.774,0.2138); rgb(221pt)=(0.9962,0.7798,0.2095); rgb(222pt)=(0.9957,0.7856,0.2053); rgb(223pt)=(0.9949,0.7915,0.2012); rgb(224pt)=(0.9938,0.7974,0.1974); rgb(225pt)=(0.9923,0.8034,0.1939); rgb(226pt)=(0.9906,0.8095,0.1906); rgb(227pt)=(0.9885,0.8156,0.1875); rgb(228pt)=(0.9861,0.8218,0.1846); rgb(229pt)=(0.9835,0.828,0.1817); rgb(230pt)=(0.9807,0.8342,0.1787); rgb(231pt)=(0.9778,0.8404,0.1757); rgb(232pt)=(0.9748,0.8467,0.1726); rgb(233pt)=(0.972,0.8529,0.1695); rgb(234pt)=(0.9694,0.8591,0.1665); rgb(235pt)=(0.9671,0.8654,0.1636); rgb(236pt)=(0.9651,0.8716,0.1608); rgb(237pt)=(0.9634,0.8778,0.1582); rgb(238pt)=(0.9619,0.884,0.1557); rgb(239pt)=(0.9608,0.8902,0.1532); rgb(240pt)=(0.9601,0.8963,0.1507); rgb(241pt)=(0.9596,0.9023,0.148); rgb(242pt)=(0.9595,0.9084,0.145); rgb(243pt)=(0.9597,0.9143,0.1418); rgb(244pt)=(0.9601,0.9203,0.1382); rgb(245pt)=(0.9608,0.9262,0.1344); rgb(246pt)=(0.9618,0.932,0.1304); rgb(247pt)=(0.9629,0.9379,0.1261); rgb(248pt)=(0.9642,0.9437,0.1216); rgb(249pt)=(0.9657,0.9494,0.1168); rgb(250pt)=(0.9674,0.9552,0.1116); rgb(251pt)=(0.9692,0.9609,0.1061); rgb(252pt)=(0.9711,0.9667,0.1001); rgb(253pt)=(0.973,0.9724,0.0938); rgb(254pt)=(0.9749,0.9782,0.0872); rgb(255pt)=(0.9769,0.9839,0.0805)},
]
\addplot [forget plot] graphics [xmin=0.01, xmax=1.98879078694818, ymin=-50.87890625, ymax=26100.87890625] {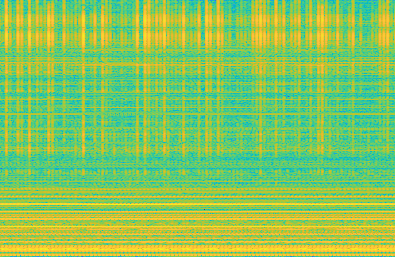};
\end{axis}
\tikzstyle{every node}=[font=\scriptsize]

\begin{axis}[%
width=0.842\fwidth,
height=0.406\fheight,
at={(0\fwidth,0\fheight)},
scale only axis,
point meta min=-33.9741477065107,
point meta max=93.9172034923056,
axis on top,
xmin=0,
xmax=80,
xlabel style={font=\color{white!15!black},font=\scriptsize},
xlabel={Modulation Frequency [Hz] $\rightarrow$},
ymin=-50.87890625,
ymax=26100.87890625,
ylabel style={font=\color{white!15!black},font=\scriptsize},
ylabel={Center Frequency [Hz] $\rightarrow$},
axis background/.style={fill=white},
legend style={legend cell align=left, align=left, draw=white!15!black},
colormap={mymap}{[1pt] rgb(0pt)=(0.2422,0.1504,0.6603); rgb(1pt)=(0.2444,0.1534,0.6728); rgb(2pt)=(0.2464,0.1569,0.6847); rgb(3pt)=(0.2484,0.1607,0.6961); rgb(4pt)=(0.2503,0.1648,0.7071); rgb(5pt)=(0.2522,0.1689,0.7179); rgb(6pt)=(0.254,0.1732,0.7286); rgb(7pt)=(0.2558,0.1773,0.7393); rgb(8pt)=(0.2576,0.1814,0.7501); rgb(9pt)=(0.2594,0.1854,0.761); rgb(11pt)=(0.2628,0.1932,0.7828); rgb(12pt)=(0.2645,0.1972,0.7937); rgb(13pt)=(0.2661,0.2011,0.8043); rgb(14pt)=(0.2676,0.2052,0.8148); rgb(15pt)=(0.2691,0.2094,0.8249); rgb(16pt)=(0.2704,0.2138,0.8346); rgb(17pt)=(0.2717,0.2184,0.8439); rgb(18pt)=(0.2729,0.2231,0.8528); rgb(19pt)=(0.274,0.228,0.8612); rgb(20pt)=(0.2749,0.233,0.8692); rgb(21pt)=(0.2758,0.2382,0.8767); rgb(22pt)=(0.2766,0.2435,0.884); rgb(23pt)=(0.2774,0.2489,0.8908); rgb(24pt)=(0.2781,0.2543,0.8973); rgb(25pt)=(0.2788,0.2598,0.9035); rgb(26pt)=(0.2794,0.2653,0.9094); rgb(27pt)=(0.2798,0.2708,0.915); rgb(28pt)=(0.2802,0.2764,0.9204); rgb(29pt)=(0.2806,0.2819,0.9255); rgb(30pt)=(0.2809,0.2875,0.9305); rgb(31pt)=(0.2811,0.293,0.9352); rgb(32pt)=(0.2813,0.2985,0.9397); rgb(33pt)=(0.2814,0.304,0.9441); rgb(34pt)=(0.2814,0.3095,0.9483); rgb(35pt)=(0.2813,0.315,0.9524); rgb(36pt)=(0.2811,0.3204,0.9563); rgb(37pt)=(0.2809,0.3259,0.96); rgb(38pt)=(0.2807,0.3313,0.9636); rgb(39pt)=(0.2803,0.3367,0.967); rgb(40pt)=(0.2798,0.3421,0.9702); rgb(41pt)=(0.2791,0.3475,0.9733); rgb(42pt)=(0.2784,0.3529,0.9763); rgb(43pt)=(0.2776,0.3583,0.9791); rgb(44pt)=(0.2766,0.3638,0.9817); rgb(45pt)=(0.2754,0.3693,0.984); rgb(46pt)=(0.2741,0.3748,0.9862); rgb(47pt)=(0.2726,0.3804,0.9881); rgb(48pt)=(0.271,0.386,0.9898); rgb(49pt)=(0.2691,0.3916,0.9912); rgb(50pt)=(0.267,0.3973,0.9924); rgb(51pt)=(0.2647,0.403,0.9935); rgb(52pt)=(0.2621,0.4088,0.9946); rgb(53pt)=(0.2591,0.4145,0.9955); rgb(54pt)=(0.2556,0.4203,0.9965); rgb(55pt)=(0.2517,0.4261,0.9974); rgb(56pt)=(0.2473,0.4319,0.9983); rgb(57pt)=(0.2424,0.4378,0.9991); rgb(58pt)=(0.2369,0.4437,0.9996); rgb(59pt)=(0.2311,0.4497,0.9995); rgb(60pt)=(0.225,0.4559,0.9985); rgb(61pt)=(0.2189,0.462,0.9968); rgb(62pt)=(0.2128,0.4682,0.9948); rgb(63pt)=(0.2066,0.4743,0.9926); rgb(64pt)=(0.2006,0.4803,0.9906); rgb(65pt)=(0.195,0.4861,0.9887); rgb(66pt)=(0.1903,0.4919,0.9867); rgb(67pt)=(0.1869,0.4975,0.9844); rgb(68pt)=(0.1847,0.503,0.9819); rgb(69pt)=(0.1831,0.5084,0.9793); rgb(70pt)=(0.1818,0.5138,0.9766); rgb(71pt)=(0.1806,0.5191,0.9738); rgb(72pt)=(0.1795,0.5244,0.9709); rgb(73pt)=(0.1785,0.5296,0.9677); rgb(74pt)=(0.1778,0.5349,0.9641); rgb(75pt)=(0.1773,0.5401,0.9602); rgb(76pt)=(0.1768,0.5452,0.956); rgb(77pt)=(0.1764,0.5504,0.9516); rgb(78pt)=(0.1755,0.5554,0.9473); rgb(79pt)=(0.174,0.5605,0.9432); rgb(80pt)=(0.1716,0.5655,0.9393); rgb(81pt)=(0.1686,0.5705,0.9357); rgb(82pt)=(0.1649,0.5755,0.9323); rgb(83pt)=(0.161,0.5805,0.9289); rgb(84pt)=(0.1573,0.5854,0.9254); rgb(85pt)=(0.154,0.5902,0.9218); rgb(86pt)=(0.1513,0.595,0.9182); rgb(87pt)=(0.1492,0.5997,0.9147); rgb(88pt)=(0.1475,0.6043,0.9113); rgb(89pt)=(0.1461,0.6089,0.908); rgb(90pt)=(0.1446,0.6135,0.905); rgb(91pt)=(0.1429,0.618,0.9022); rgb(92pt)=(0.1408,0.6226,0.8998); rgb(93pt)=(0.1383,0.6272,0.8975); rgb(94pt)=(0.1354,0.6317,0.8953); rgb(95pt)=(0.1321,0.6363,0.8932); rgb(96pt)=(0.1288,0.6408,0.891); rgb(97pt)=(0.1253,0.6453,0.8887); rgb(98pt)=(0.1219,0.6497,0.8862); rgb(99pt)=(0.1185,0.6541,0.8834); rgb(100pt)=(0.1152,0.6584,0.8804); rgb(101pt)=(0.1119,0.6627,0.877); rgb(102pt)=(0.1085,0.6669,0.8734); rgb(103pt)=(0.1048,0.671,0.8695); rgb(104pt)=(0.1009,0.675,0.8653); rgb(105pt)=(0.0964,0.6789,0.8609); rgb(106pt)=(0.0914,0.6828,0.8562); rgb(107pt)=(0.0855,0.6865,0.8513); rgb(108pt)=(0.0789,0.6902,0.8462); rgb(109pt)=(0.0713,0.6938,0.8409); rgb(110pt)=(0.0628,0.6972,0.8355); rgb(111pt)=(0.0535,0.7006,0.8299); rgb(112pt)=(0.0433,0.7039,0.8242); rgb(113pt)=(0.0328,0.7071,0.8183); rgb(114pt)=(0.0234,0.7103,0.8124); rgb(115pt)=(0.0155,0.7133,0.8064); rgb(116pt)=(0.0091,0.7163,0.8003); rgb(117pt)=(0.0046,0.7192,0.7941); rgb(118pt)=(0.0019,0.722,0.7878); rgb(119pt)=(0.0009,0.7248,0.7815); rgb(120pt)=(0.0018,0.7275,0.7752); rgb(121pt)=(0.0046,0.7301,0.7688); rgb(122pt)=(0.0094,0.7327,0.7623); rgb(123pt)=(0.0162,0.7352,0.7558); rgb(124pt)=(0.0253,0.7376,0.7492); rgb(125pt)=(0.0369,0.74,0.7426); rgb(126pt)=(0.0504,0.7423,0.7359); rgb(127pt)=(0.0638,0.7446,0.7292); rgb(128pt)=(0.077,0.7468,0.7224); rgb(129pt)=(0.0899,0.7489,0.7156); rgb(130pt)=(0.1023,0.751,0.7088); rgb(131pt)=(0.1141,0.7531,0.7019); rgb(132pt)=(0.1252,0.7552,0.695); rgb(133pt)=(0.1354,0.7572,0.6881); rgb(134pt)=(0.1448,0.7593,0.6812); rgb(135pt)=(0.1532,0.7614,0.6741); rgb(136pt)=(0.1609,0.7635,0.6671); rgb(137pt)=(0.1678,0.7656,0.6599); rgb(138pt)=(0.1741,0.7678,0.6527); rgb(139pt)=(0.1799,0.7699,0.6454); rgb(140pt)=(0.1853,0.7721,0.6379); rgb(141pt)=(0.1905,0.7743,0.6303); rgb(142pt)=(0.1954,0.7765,0.6225); rgb(143pt)=(0.2003,0.7787,0.6146); rgb(144pt)=(0.2061,0.7808,0.6065); rgb(145pt)=(0.2118,0.7828,0.5983); rgb(146pt)=(0.2178,0.7849,0.5899); rgb(147pt)=(0.2244,0.7869,0.5813); rgb(148pt)=(0.2318,0.7887,0.5725); rgb(149pt)=(0.2401,0.7905,0.5636); rgb(150pt)=(0.2491,0.7922,0.5546); rgb(151pt)=(0.2589,0.7937,0.5454); rgb(152pt)=(0.2695,0.7951,0.536); rgb(153pt)=(0.2809,0.7964,0.5266); rgb(154pt)=(0.2929,0.7975,0.517); rgb(155pt)=(0.3052,0.7985,0.5074); rgb(156pt)=(0.3176,0.7994,0.4975); rgb(157pt)=(0.3301,0.8002,0.4876); rgb(158pt)=(0.3424,0.8009,0.4774); rgb(159pt)=(0.3548,0.8016,0.4669); rgb(160pt)=(0.3671,0.8021,0.4563); rgb(161pt)=(0.3795,0.8026,0.4454); rgb(162pt)=(0.3921,0.8029,0.4344); rgb(163pt)=(0.405,0.8031,0.4233); rgb(164pt)=(0.4184,0.803,0.4122); rgb(165pt)=(0.4322,0.8028,0.4013); rgb(166pt)=(0.4463,0.8024,0.3904); rgb(167pt)=(0.4608,0.8018,0.3797); rgb(168pt)=(0.4753,0.8011,0.3691); rgb(169pt)=(0.4899,0.8002,0.3586); rgb(170pt)=(0.5044,0.7993,0.348); rgb(171pt)=(0.5187,0.7982,0.3374); rgb(172pt)=(0.5329,0.797,0.3267); rgb(173pt)=(0.547,0.7957,0.3159); rgb(175pt)=(0.5748,0.7929,0.2941); rgb(176pt)=(0.5886,0.7913,0.2833); rgb(177pt)=(0.6024,0.7896,0.2726); rgb(178pt)=(0.6161,0.7878,0.2622); rgb(179pt)=(0.6297,0.7859,0.2521); rgb(180pt)=(0.6433,0.7839,0.2423); rgb(181pt)=(0.6567,0.7818,0.2329); rgb(182pt)=(0.6701,0.7796,0.2239); rgb(183pt)=(0.6833,0.7773,0.2155); rgb(184pt)=(0.6963,0.775,0.2075); rgb(185pt)=(0.7091,0.7727,0.1998); rgb(186pt)=(0.7218,0.7703,0.1924); rgb(187pt)=(0.7344,0.7679,0.1852); rgb(188pt)=(0.7468,0.7654,0.1782); rgb(189pt)=(0.759,0.7629,0.1717); rgb(190pt)=(0.771,0.7604,0.1658); rgb(191pt)=(0.7829,0.7579,0.1608); rgb(192pt)=(0.7945,0.7554,0.157); rgb(193pt)=(0.806,0.7529,0.1546); rgb(194pt)=(0.8172,0.7505,0.1535); rgb(195pt)=(0.8281,0.7481,0.1536); rgb(196pt)=(0.8389,0.7457,0.1546); rgb(197pt)=(0.8495,0.7435,0.1564); rgb(198pt)=(0.86,0.7413,0.1587); rgb(199pt)=(0.8703,0.7392,0.1615); rgb(200pt)=(0.8804,0.7372,0.165); rgb(201pt)=(0.8903,0.7353,0.1695); rgb(202pt)=(0.9,0.7336,0.1749); rgb(203pt)=(0.9093,0.7321,0.1815); rgb(204pt)=(0.9184,0.7308,0.189); rgb(205pt)=(0.9272,0.7298,0.1973); rgb(206pt)=(0.9357,0.729,0.2061); rgb(207pt)=(0.944,0.7285,0.2151); rgb(208pt)=(0.9523,0.7284,0.2237); rgb(209pt)=(0.9606,0.7285,0.2312); rgb(210pt)=(0.9689,0.7292,0.2373); rgb(211pt)=(0.977,0.7304,0.2418); rgb(212pt)=(0.9842,0.733,0.2446); rgb(213pt)=(0.99,0.7365,0.2429); rgb(214pt)=(0.9946,0.7407,0.2394); rgb(215pt)=(0.9966,0.7458,0.2351); rgb(216pt)=(0.9971,0.7513,0.2309); rgb(217pt)=(0.9972,0.7569,0.2267); rgb(218pt)=(0.9971,0.7626,0.2224); rgb(219pt)=(0.9969,0.7683,0.2181); rgb(220pt)=(0.9966,0.774,0.2138); rgb(221pt)=(0.9962,0.7798,0.2095); rgb(222pt)=(0.9957,0.7856,0.2053); rgb(223pt)=(0.9949,0.7915,0.2012); rgb(224pt)=(0.9938,0.7974,0.1974); rgb(225pt)=(0.9923,0.8034,0.1939); rgb(226pt)=(0.9906,0.8095,0.1906); rgb(227pt)=(0.9885,0.8156,0.1875); rgb(228pt)=(0.9861,0.8218,0.1846); rgb(229pt)=(0.9835,0.828,0.1817); rgb(230pt)=(0.9807,0.8342,0.1787); rgb(231pt)=(0.9778,0.8404,0.1757); rgb(232pt)=(0.9748,0.8467,0.1726); rgb(233pt)=(0.972,0.8529,0.1695); rgb(234pt)=(0.9694,0.8591,0.1665); rgb(235pt)=(0.9671,0.8654,0.1636); rgb(236pt)=(0.9651,0.8716,0.1608); rgb(237pt)=(0.9634,0.8778,0.1582); rgb(238pt)=(0.9619,0.884,0.1557); rgb(239pt)=(0.9608,0.8902,0.1532); rgb(240pt)=(0.9601,0.8963,0.1507); rgb(241pt)=(0.9596,0.9023,0.148); rgb(242pt)=(0.9595,0.9084,0.145); rgb(243pt)=(0.9597,0.9143,0.1418); rgb(244pt)=(0.9601,0.9203,0.1382); rgb(245pt)=(0.9608,0.9262,0.1344); rgb(246pt)=(0.9618,0.932,0.1304); rgb(247pt)=(0.9629,0.9379,0.1261); rgb(248pt)=(0.9642,0.9437,0.1216); rgb(249pt)=(0.9657,0.9494,0.1168); rgb(250pt)=(0.9674,0.9552,0.1116); rgb(251pt)=(0.9692,0.9609,0.1061); rgb(252pt)=(0.9711,0.9667,0.1001); rgb(253pt)=(0.973,0.9724,0.0938); rgb(254pt)=(0.9749,0.9782,0.0872); rgb(255pt)=(0.9769,0.9839,0.0805)},
]
\addplot [forget plot] graphics [xmin=-0.390625, xmax=100.390625, ymin=-50.87890625, ymax=26100.87890625] {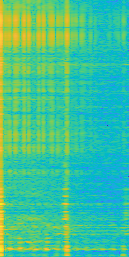};
\draw[color=red,dashed,ultra thick]  (0.5,200) rectangle (80,260); 
 \end{axis}
\end{tikzpicture}%

%% file: figures/ams_healthy.tex
%
%
\begin{tikzpicture}
\tikzstyle{every node}=[font=\scriptsize]

\begin{axis}[%
width=0.842\fwidth,
height=0.406\fheight,
at={(0\fwidth,0.594\fheight)},
scale only axis,
point meta min=-56.4935557892843,
point meta max=60.2511067873325,
axis on top,
xmin=0.01,
xmax=1.98879078694818,
xlabel style={font=\color{white!15!black},font=\scriptsize},
xlabel={Time [s] $\rightarrow$},
ymin=-50.87890625,
ymax=26.1e3,
ylabel style={font=\color{white!15!black},font=\scriptsize},
ylabel={Frequency [Hz] $\rightarrow$},
axis background/.style={fill=white},
legend style={legend cell align=left, align=left, draw=white!15!black},
colormap={mymap}{[1pt] rgb(0pt)=(0.2422,0.1504,0.6603); rgb(1pt)=(0.2444,0.1534,0.6728); rgb(2pt)=(0.2464,0.1569,0.6847); rgb(3pt)=(0.2484,0.1607,0.6961); rgb(4pt)=(0.2503,0.1648,0.7071); rgb(5pt)=(0.2522,0.1689,0.7179); rgb(6pt)=(0.254,0.1732,0.7286); rgb(7pt)=(0.2558,0.1773,0.7393); rgb(8pt)=(0.2576,0.1814,0.7501); rgb(9pt)=(0.2594,0.1854,0.761); rgb(11pt)=(0.2628,0.1932,0.7828); rgb(12pt)=(0.2645,0.1972,0.7937); rgb(13pt)=(0.2661,0.2011,0.8043); rgb(14pt)=(0.2676,0.2052,0.8148); rgb(15pt)=(0.2691,0.2094,0.8249); rgb(16pt)=(0.2704,0.2138,0.8346); rgb(17pt)=(0.2717,0.2184,0.8439); rgb(18pt)=(0.2729,0.2231,0.8528); rgb(19pt)=(0.274,0.228,0.8612); rgb(20pt)=(0.2749,0.233,0.8692); rgb(21pt)=(0.2758,0.2382,0.8767); rgb(22pt)=(0.2766,0.2435,0.884); rgb(23pt)=(0.2774,0.2489,0.8908); rgb(24pt)=(0.2781,0.2543,0.8973); rgb(25pt)=(0.2788,0.2598,0.9035); rgb(26pt)=(0.2794,0.2653,0.9094); rgb(27pt)=(0.2798,0.2708,0.915); rgb(28pt)=(0.2802,0.2764,0.9204); rgb(29pt)=(0.2806,0.2819,0.9255); rgb(30pt)=(0.2809,0.2875,0.9305); rgb(31pt)=(0.2811,0.293,0.9352); rgb(32pt)=(0.2813,0.2985,0.9397); rgb(33pt)=(0.2814,0.304,0.9441); rgb(34pt)=(0.2814,0.3095,0.9483); rgb(35pt)=(0.2813,0.315,0.9524); rgb(36pt)=(0.2811,0.3204,0.9563); rgb(37pt)=(0.2809,0.3259,0.96); rgb(38pt)=(0.2807,0.3313,0.9636); rgb(39pt)=(0.2803,0.3367,0.967); rgb(40pt)=(0.2798,0.3421,0.9702); rgb(41pt)=(0.2791,0.3475,0.9733); rgb(42pt)=(0.2784,0.3529,0.9763); rgb(43pt)=(0.2776,0.3583,0.9791); rgb(44pt)=(0.2766,0.3638,0.9817); rgb(45pt)=(0.2754,0.3693,0.984); rgb(46pt)=(0.2741,0.3748,0.9862); rgb(47pt)=(0.2726,0.3804,0.9881); rgb(48pt)=(0.271,0.386,0.9898); rgb(49pt)=(0.2691,0.3916,0.9912); rgb(50pt)=(0.267,0.3973,0.9924); rgb(51pt)=(0.2647,0.403,0.9935); rgb(52pt)=(0.2621,0.4088,0.9946); rgb(53pt)=(0.2591,0.4145,0.9955); rgb(54pt)=(0.2556,0.4203,0.9965); rgb(55pt)=(0.2517,0.4261,0.9974); rgb(56pt)=(0.2473,0.4319,0.9983); rgb(57pt)=(0.2424,0.4378,0.9991); rgb(58pt)=(0.2369,0.4437,0.9996); rgb(59pt)=(0.2311,0.4497,0.9995); rgb(60pt)=(0.225,0.4559,0.9985); rgb(61pt)=(0.2189,0.462,0.9968); rgb(62pt)=(0.2128,0.4682,0.9948); rgb(63pt)=(0.2066,0.4743,0.9926); rgb(64pt)=(0.2006,0.4803,0.9906); rgb(65pt)=(0.195,0.4861,0.9887); rgb(66pt)=(0.1903,0.4919,0.9867); rgb(67pt)=(0.1869,0.4975,0.9844); rgb(68pt)=(0.1847,0.503,0.9819); rgb(69pt)=(0.1831,0.5084,0.9793); rgb(70pt)=(0.1818,0.5138,0.9766); rgb(71pt)=(0.1806,0.5191,0.9738); rgb(72pt)=(0.1795,0.5244,0.9709); rgb(73pt)=(0.1785,0.5296,0.9677); rgb(74pt)=(0.1778,0.5349,0.9641); rgb(75pt)=(0.1773,0.5401,0.9602); rgb(76pt)=(0.1768,0.5452,0.956); rgb(77pt)=(0.1764,0.5504,0.9516); rgb(78pt)=(0.1755,0.5554,0.9473); rgb(79pt)=(0.174,0.5605,0.9432); rgb(80pt)=(0.1716,0.5655,0.9393); rgb(81pt)=(0.1686,0.5705,0.9357); rgb(82pt)=(0.1649,0.5755,0.9323); rgb(83pt)=(0.161,0.5805,0.9289); rgb(84pt)=(0.1573,0.5854,0.9254); rgb(85pt)=(0.154,0.5902,0.9218); rgb(86pt)=(0.1513,0.595,0.9182); rgb(87pt)=(0.1492,0.5997,0.9147); rgb(88pt)=(0.1475,0.6043,0.9113); rgb(89pt)=(0.1461,0.6089,0.908); rgb(90pt)=(0.1446,0.6135,0.905); rgb(91pt)=(0.1429,0.618,0.9022); rgb(92pt)=(0.1408,0.6226,0.8998); rgb(93pt)=(0.1383,0.6272,0.8975); rgb(94pt)=(0.1354,0.6317,0.8953); rgb(95pt)=(0.1321,0.6363,0.8932); rgb(96pt)=(0.1288,0.6408,0.891); rgb(97pt)=(0.1253,0.6453,0.8887); rgb(98pt)=(0.1219,0.6497,0.8862); rgb(99pt)=(0.1185,0.6541,0.8834); rgb(100pt)=(0.1152,0.6584,0.8804); rgb(101pt)=(0.1119,0.6627,0.877); rgb(102pt)=(0.1085,0.6669,0.8734); rgb(103pt)=(0.1048,0.671,0.8695); rgb(104pt)=(0.1009,0.675,0.8653); rgb(105pt)=(0.0964,0.6789,0.8609); rgb(106pt)=(0.0914,0.6828,0.8562); rgb(107pt)=(0.0855,0.6865,0.8513); rgb(108pt)=(0.0789,0.6902,0.8462); rgb(109pt)=(0.0713,0.6938,0.8409); rgb(110pt)=(0.0628,0.6972,0.8355); rgb(111pt)=(0.0535,0.7006,0.8299); rgb(112pt)=(0.0433,0.7039,0.8242); rgb(113pt)=(0.0328,0.7071,0.8183); rgb(114pt)=(0.0234,0.7103,0.8124); rgb(115pt)=(0.0155,0.7133,0.8064); rgb(116pt)=(0.0091,0.7163,0.8003); rgb(117pt)=(0.0046,0.7192,0.7941); rgb(118pt)=(0.0019,0.722,0.7878); rgb(119pt)=(0.0009,0.7248,0.7815); rgb(120pt)=(0.0018,0.7275,0.7752); rgb(121pt)=(0.0046,0.7301,0.7688); rgb(122pt)=(0.0094,0.7327,0.7623); rgb(123pt)=(0.0162,0.7352,0.7558); rgb(124pt)=(0.0253,0.7376,0.7492); rgb(125pt)=(0.0369,0.74,0.7426); rgb(126pt)=(0.0504,0.7423,0.7359); rgb(127pt)=(0.0638,0.7446,0.7292); rgb(128pt)=(0.077,0.7468,0.7224); rgb(129pt)=(0.0899,0.7489,0.7156); rgb(130pt)=(0.1023,0.751,0.7088); rgb(131pt)=(0.1141,0.7531,0.7019); rgb(132pt)=(0.1252,0.7552,0.695); rgb(133pt)=(0.1354,0.7572,0.6881); rgb(134pt)=(0.1448,0.7593,0.6812); rgb(135pt)=(0.1532,0.7614,0.6741); rgb(136pt)=(0.1609,0.7635,0.6671); rgb(137pt)=(0.1678,0.7656,0.6599); rgb(138pt)=(0.1741,0.7678,0.6527); rgb(139pt)=(0.1799,0.7699,0.6454); rgb(140pt)=(0.1853,0.7721,0.6379); rgb(141pt)=(0.1905,0.7743,0.6303); rgb(142pt)=(0.1954,0.7765,0.6225); rgb(143pt)=(0.2003,0.7787,0.6146); rgb(144pt)=(0.2061,0.7808,0.6065); rgb(145pt)=(0.2118,0.7828,0.5983); rgb(146pt)=(0.2178,0.7849,0.5899); rgb(147pt)=(0.2244,0.7869,0.5813); rgb(148pt)=(0.2318,0.7887,0.5725); rgb(149pt)=(0.2401,0.7905,0.5636); rgb(150pt)=(0.2491,0.7922,0.5546); rgb(151pt)=(0.2589,0.7937,0.5454); rgb(152pt)=(0.2695,0.7951,0.536); rgb(153pt)=(0.2809,0.7964,0.5266); rgb(154pt)=(0.2929,0.7975,0.517); rgb(155pt)=(0.3052,0.7985,0.5074); rgb(156pt)=(0.3176,0.7994,0.4975); rgb(157pt)=(0.3301,0.8002,0.4876); rgb(158pt)=(0.3424,0.8009,0.4774); rgb(159pt)=(0.3548,0.8016,0.4669); rgb(160pt)=(0.3671,0.8021,0.4563); rgb(161pt)=(0.3795,0.8026,0.4454); rgb(162pt)=(0.3921,0.8029,0.4344); rgb(163pt)=(0.405,0.8031,0.4233); rgb(164pt)=(0.4184,0.803,0.4122); rgb(165pt)=(0.4322,0.8028,0.4013); rgb(166pt)=(0.4463,0.8024,0.3904); rgb(167pt)=(0.4608,0.8018,0.3797); rgb(168pt)=(0.4753,0.8011,0.3691); rgb(169pt)=(0.4899,0.8002,0.3586); rgb(170pt)=(0.5044,0.7993,0.348); rgb(171pt)=(0.5187,0.7982,0.3374); rgb(172pt)=(0.5329,0.797,0.3267); rgb(173pt)=(0.547,0.7957,0.3159); rgb(175pt)=(0.5748,0.7929,0.2941); rgb(176pt)=(0.5886,0.7913,0.2833); rgb(177pt)=(0.6024,0.7896,0.2726); rgb(178pt)=(0.6161,0.7878,0.2622); rgb(179pt)=(0.6297,0.7859,0.2521); rgb(180pt)=(0.6433,0.7839,0.2423); rgb(181pt)=(0.6567,0.7818,0.2329); rgb(182pt)=(0.6701,0.7796,0.2239); rgb(183pt)=(0.6833,0.7773,0.2155); rgb(184pt)=(0.6963,0.775,0.2075); rgb(185pt)=(0.7091,0.7727,0.1998); rgb(186pt)=(0.7218,0.7703,0.1924); rgb(187pt)=(0.7344,0.7679,0.1852); rgb(188pt)=(0.7468,0.7654,0.1782); rgb(189pt)=(0.759,0.7629,0.1717); rgb(190pt)=(0.771,0.7604,0.1658); rgb(191pt)=(0.7829,0.7579,0.1608); rgb(192pt)=(0.7945,0.7554,0.157); rgb(193pt)=(0.806,0.7529,0.1546); rgb(194pt)=(0.8172,0.7505,0.1535); rgb(195pt)=(0.8281,0.7481,0.1536); rgb(196pt)=(0.8389,0.7457,0.1546); rgb(197pt)=(0.8495,0.7435,0.1564); rgb(198pt)=(0.86,0.7413,0.1587); rgb(199pt)=(0.8703,0.7392,0.1615); rgb(200pt)=(0.8804,0.7372,0.165); rgb(201pt)=(0.8903,0.7353,0.1695); rgb(202pt)=(0.9,0.7336,0.1749); rgb(203pt)=(0.9093,0.7321,0.1815); rgb(204pt)=(0.9184,0.7308,0.189); rgb(205pt)=(0.9272,0.7298,0.1973); rgb(206pt)=(0.9357,0.729,0.2061); rgb(207pt)=(0.944,0.7285,0.2151); rgb(208pt)=(0.9523,0.7284,0.2237); rgb(209pt)=(0.9606,0.7285,0.2312); rgb(210pt)=(0.9689,0.7292,0.2373); rgb(211pt)=(0.977,0.7304,0.2418); rgb(212pt)=(0.9842,0.733,0.2446); rgb(213pt)=(0.99,0.7365,0.2429); rgb(214pt)=(0.9946,0.7407,0.2394); rgb(215pt)=(0.9966,0.7458,0.2351); rgb(216pt)=(0.9971,0.7513,0.2309); rgb(217pt)=(0.9972,0.7569,0.2267); rgb(218pt)=(0.9971,0.7626,0.2224); rgb(219pt)=(0.9969,0.7683,0.2181); rgb(220pt)=(0.9966,0.774,0.2138); rgb(221pt)=(0.9962,0.7798,0.2095); rgb(222pt)=(0.9957,0.7856,0.2053); rgb(223pt)=(0.9949,0.7915,0.2012); rgb(224pt)=(0.9938,0.7974,0.1974); rgb(225pt)=(0.9923,0.8034,0.1939); rgb(226pt)=(0.9906,0.8095,0.1906); rgb(227pt)=(0.9885,0.8156,0.1875); rgb(228pt)=(0.9861,0.8218,0.1846); rgb(229pt)=(0.9835,0.828,0.1817); rgb(230pt)=(0.9807,0.8342,0.1787); rgb(231pt)=(0.9778,0.8404,0.1757); rgb(232pt)=(0.9748,0.8467,0.1726); rgb(233pt)=(0.972,0.8529,0.1695); rgb(234pt)=(0.9694,0.8591,0.1665); rgb(235pt)=(0.9671,0.8654,0.1636); rgb(236pt)=(0.9651,0.8716,0.1608); rgb(237pt)=(0.9634,0.8778,0.1582); rgb(238pt)=(0.9619,0.884,0.1557); rgb(239pt)=(0.9608,0.8902,0.1532); rgb(240pt)=(0.9601,0.8963,0.1507); rgb(241pt)=(0.9596,0.9023,0.148); rgb(242pt)=(0.9595,0.9084,0.145); rgb(243pt)=(0.9597,0.9143,0.1418); rgb(244pt)=(0.9601,0.9203,0.1382); rgb(245pt)=(0.9608,0.9262,0.1344); rgb(246pt)=(0.9618,0.932,0.1304); rgb(247pt)=(0.9629,0.9379,0.1261); rgb(248pt)=(0.9642,0.9437,0.1216); rgb(249pt)=(0.9657,0.9494,0.1168); rgb(250pt)=(0.9674,0.9552,0.1116); rgb(251pt)=(0.9692,0.9609,0.1061); rgb(252pt)=(0.9711,0.9667,0.1001); rgb(253pt)=(0.973,0.9724,0.0938); rgb(254pt)=(0.9749,0.9782,0.0872); rgb(255pt)=(0.9769,0.9839,0.0805)},
]
\addplot [forget plot] graphics [xmin=0.01, xmax=1.98879078694818, ymin=-50.87890625, ymax=26100.87890625] {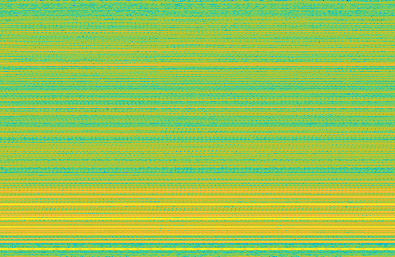};
\end{axis}
\tikzstyle{every node}=[font=\scriptsize]

\begin{axis}[%
width=0.842\fwidth,
height=0.406\fheight,
at={(0\fwidth,0\fheight)},
scale only axis,
point meta min=-27.3948456897245,
point meta max=95.7109742674355,
axis on top,
xmin=0,
xmax=80,
xlabel style={font=\color{white!15!black},font=\scriptsize},
xlabel={Modulation Frequency [Hz] $\rightarrow$},
ymin=-50.87890625,
ymax=26100.87890625,
ylabel style={font=\color{white!15!black},font=\scriptsize},
ylabel={Center Frequency [Hz] $\rightarrow$},
axis background/.style={fill=white},
legend style={legend cell align=left, align=left, draw=white!15!black},
colormap={mymap}{[1pt] rgb(0pt)=(0.2422,0.1504,0.6603); rgb(1pt)=(0.2444,0.1534,0.6728); rgb(2pt)=(0.2464,0.1569,0.6847); rgb(3pt)=(0.2484,0.1607,0.6961); rgb(4pt)=(0.2503,0.1648,0.7071); rgb(5pt)=(0.2522,0.1689,0.7179); rgb(6pt)=(0.254,0.1732,0.7286); rgb(7pt)=(0.2558,0.1773,0.7393); rgb(8pt)=(0.2576,0.1814,0.7501); rgb(9pt)=(0.2594,0.1854,0.761); rgb(11pt)=(0.2628,0.1932,0.7828); rgb(12pt)=(0.2645,0.1972,0.7937); rgb(13pt)=(0.2661,0.2011,0.8043); rgb(14pt)=(0.2676,0.2052,0.8148); rgb(15pt)=(0.2691,0.2094,0.8249); rgb(16pt)=(0.2704,0.2138,0.8346); rgb(17pt)=(0.2717,0.2184,0.8439); rgb(18pt)=(0.2729,0.2231,0.8528); rgb(19pt)=(0.274,0.228,0.8612); rgb(20pt)=(0.2749,0.233,0.8692); rgb(21pt)=(0.2758,0.2382,0.8767); rgb(22pt)=(0.2766,0.2435,0.884); rgb(23pt)=(0.2774,0.2489,0.8908); rgb(24pt)=(0.2781,0.2543,0.8973); rgb(25pt)=(0.2788,0.2598,0.9035); rgb(26pt)=(0.2794,0.2653,0.9094); rgb(27pt)=(0.2798,0.2708,0.915); rgb(28pt)=(0.2802,0.2764,0.9204); rgb(29pt)=(0.2806,0.2819,0.9255); rgb(30pt)=(0.2809,0.2875,0.9305); rgb(31pt)=(0.2811,0.293,0.9352); rgb(32pt)=(0.2813,0.2985,0.9397); rgb(33pt)=(0.2814,0.304,0.9441); rgb(34pt)=(0.2814,0.3095,0.9483); rgb(35pt)=(0.2813,0.315,0.9524); rgb(36pt)=(0.2811,0.3204,0.9563); rgb(37pt)=(0.2809,0.3259,0.96); rgb(38pt)=(0.2807,0.3313,0.9636); rgb(39pt)=(0.2803,0.3367,0.967); rgb(40pt)=(0.2798,0.3421,0.9702); rgb(41pt)=(0.2791,0.3475,0.9733); rgb(42pt)=(0.2784,0.3529,0.9763); rgb(43pt)=(0.2776,0.3583,0.9791); rgb(44pt)=(0.2766,0.3638,0.9817); rgb(45pt)=(0.2754,0.3693,0.984); rgb(46pt)=(0.2741,0.3748,0.9862); rgb(47pt)=(0.2726,0.3804,0.9881); rgb(48pt)=(0.271,0.386,0.9898); rgb(49pt)=(0.2691,0.3916,0.9912); rgb(50pt)=(0.267,0.3973,0.9924); rgb(51pt)=(0.2647,0.403,0.9935); rgb(52pt)=(0.2621,0.4088,0.9946); rgb(53pt)=(0.2591,0.4145,0.9955); rgb(54pt)=(0.2556,0.4203,0.9965); rgb(55pt)=(0.2517,0.4261,0.9974); rgb(56pt)=(0.2473,0.4319,0.9983); rgb(57pt)=(0.2424,0.4378,0.9991); rgb(58pt)=(0.2369,0.4437,0.9996); rgb(59pt)=(0.2311,0.4497,0.9995); rgb(60pt)=(0.225,0.4559,0.9985); rgb(61pt)=(0.2189,0.462,0.9968); rgb(62pt)=(0.2128,0.4682,0.9948); rgb(63pt)=(0.2066,0.4743,0.9926); rgb(64pt)=(0.2006,0.4803,0.9906); rgb(65pt)=(0.195,0.4861,0.9887); rgb(66pt)=(0.1903,0.4919,0.9867); rgb(67pt)=(0.1869,0.4975,0.9844); rgb(68pt)=(0.1847,0.503,0.9819); rgb(69pt)=(0.1831,0.5084,0.9793); rgb(70pt)=(0.1818,0.5138,0.9766); rgb(71pt)=(0.1806,0.5191,0.9738); rgb(72pt)=(0.1795,0.5244,0.9709); rgb(73pt)=(0.1785,0.5296,0.9677); rgb(74pt)=(0.1778,0.5349,0.9641); rgb(75pt)=(0.1773,0.5401,0.9602); rgb(76pt)=(0.1768,0.5452,0.956); rgb(77pt)=(0.1764,0.5504,0.9516); rgb(78pt)=(0.1755,0.5554,0.9473); rgb(79pt)=(0.174,0.5605,0.9432); rgb(80pt)=(0.1716,0.5655,0.9393); rgb(81pt)=(0.1686,0.5705,0.9357); rgb(82pt)=(0.1649,0.5755,0.9323); rgb(83pt)=(0.161,0.5805,0.9289); rgb(84pt)=(0.1573,0.5854,0.9254); rgb(85pt)=(0.154,0.5902,0.9218); rgb(86pt)=(0.1513,0.595,0.9182); rgb(87pt)=(0.1492,0.5997,0.9147); rgb(88pt)=(0.1475,0.6043,0.9113); rgb(89pt)=(0.1461,0.6089,0.908); rgb(90pt)=(0.1446,0.6135,0.905); rgb(91pt)=(0.1429,0.618,0.9022); rgb(92pt)=(0.1408,0.6226,0.8998); rgb(93pt)=(0.1383,0.6272,0.8975); rgb(94pt)=(0.1354,0.6317,0.8953); rgb(95pt)=(0.1321,0.6363,0.8932); rgb(96pt)=(0.1288,0.6408,0.891); rgb(97pt)=(0.1253,0.6453,0.8887); rgb(98pt)=(0.1219,0.6497,0.8862); rgb(99pt)=(0.1185,0.6541,0.8834); rgb(100pt)=(0.1152,0.6584,0.8804); rgb(101pt)=(0.1119,0.6627,0.877); rgb(102pt)=(0.1085,0.6669,0.8734); rgb(103pt)=(0.1048,0.671,0.8695); rgb(104pt)=(0.1009,0.675,0.8653); rgb(105pt)=(0.0964,0.6789,0.8609); rgb(106pt)=(0.0914,0.6828,0.8562); rgb(107pt)=(0.0855,0.6865,0.8513); rgb(108pt)=(0.0789,0.6902,0.8462); rgb(109pt)=(0.0713,0.6938,0.8409); rgb(110pt)=(0.0628,0.6972,0.8355); rgb(111pt)=(0.0535,0.7006,0.8299); rgb(112pt)=(0.0433,0.7039,0.8242); rgb(113pt)=(0.0328,0.7071,0.8183); rgb(114pt)=(0.0234,0.7103,0.8124); rgb(115pt)=(0.0155,0.7133,0.8064); rgb(116pt)=(0.0091,0.7163,0.8003); rgb(117pt)=(0.0046,0.7192,0.7941); rgb(118pt)=(0.0019,0.722,0.7878); rgb(119pt)=(0.0009,0.7248,0.7815); rgb(120pt)=(0.0018,0.7275,0.7752); rgb(121pt)=(0.0046,0.7301,0.7688); rgb(122pt)=(0.0094,0.7327,0.7623); rgb(123pt)=(0.0162,0.7352,0.7558); rgb(124pt)=(0.0253,0.7376,0.7492); rgb(125pt)=(0.0369,0.74,0.7426); rgb(126pt)=(0.0504,0.7423,0.7359); rgb(127pt)=(0.0638,0.7446,0.7292); rgb(128pt)=(0.077,0.7468,0.7224); rgb(129pt)=(0.0899,0.7489,0.7156); rgb(130pt)=(0.1023,0.751,0.7088); rgb(131pt)=(0.1141,0.7531,0.7019); rgb(132pt)=(0.1252,0.7552,0.695); rgb(133pt)=(0.1354,0.7572,0.6881); rgb(134pt)=(0.1448,0.7593,0.6812); rgb(135pt)=(0.1532,0.7614,0.6741); rgb(136pt)=(0.1609,0.7635,0.6671); rgb(137pt)=(0.1678,0.7656,0.6599); rgb(138pt)=(0.1741,0.7678,0.6527); rgb(139pt)=(0.1799,0.7699,0.6454); rgb(140pt)=(0.1853,0.7721,0.6379); rgb(141pt)=(0.1905,0.7743,0.6303); rgb(142pt)=(0.1954,0.7765,0.6225); rgb(143pt)=(0.2003,0.7787,0.6146); rgb(144pt)=(0.2061,0.7808,0.6065); rgb(145pt)=(0.2118,0.7828,0.5983); rgb(146pt)=(0.2178,0.7849,0.5899); rgb(147pt)=(0.2244,0.7869,0.5813); rgb(148pt)=(0.2318,0.7887,0.5725); rgb(149pt)=(0.2401,0.7905,0.5636); rgb(150pt)=(0.2491,0.7922,0.5546); rgb(151pt)=(0.2589,0.7937,0.5454); rgb(152pt)=(0.2695,0.7951,0.536); rgb(153pt)=(0.2809,0.7964,0.5266); rgb(154pt)=(0.2929,0.7975,0.517); rgb(155pt)=(0.3052,0.7985,0.5074); rgb(156pt)=(0.3176,0.7994,0.4975); rgb(157pt)=(0.3301,0.8002,0.4876); rgb(158pt)=(0.3424,0.8009,0.4774); rgb(159pt)=(0.3548,0.8016,0.4669); rgb(160pt)=(0.3671,0.8021,0.4563); rgb(161pt)=(0.3795,0.8026,0.4454); rgb(162pt)=(0.3921,0.8029,0.4344); rgb(163pt)=(0.405,0.8031,0.4233); rgb(164pt)=(0.4184,0.803,0.4122); rgb(165pt)=(0.4322,0.8028,0.4013); rgb(166pt)=(0.4463,0.8024,0.3904); rgb(167pt)=(0.4608,0.8018,0.3797); rgb(168pt)=(0.4753,0.8011,0.3691); rgb(169pt)=(0.4899,0.8002,0.3586); rgb(170pt)=(0.5044,0.7993,0.348); rgb(171pt)=(0.5187,0.7982,0.3374); rgb(172pt)=(0.5329,0.797,0.3267); rgb(173pt)=(0.547,0.7957,0.3159); rgb(175pt)=(0.5748,0.7929,0.2941); rgb(176pt)=(0.5886,0.7913,0.2833); rgb(177pt)=(0.6024,0.7896,0.2726); rgb(178pt)=(0.6161,0.7878,0.2622); rgb(179pt)=(0.6297,0.7859,0.2521); rgb(180pt)=(0.6433,0.7839,0.2423); rgb(181pt)=(0.6567,0.7818,0.2329); rgb(182pt)=(0.6701,0.7796,0.2239); rgb(183pt)=(0.6833,0.7773,0.2155); rgb(184pt)=(0.6963,0.775,0.2075); rgb(185pt)=(0.7091,0.7727,0.1998); rgb(186pt)=(0.7218,0.7703,0.1924); rgb(187pt)=(0.7344,0.7679,0.1852); rgb(188pt)=(0.7468,0.7654,0.1782); rgb(189pt)=(0.759,0.7629,0.1717); rgb(190pt)=(0.771,0.7604,0.1658); rgb(191pt)=(0.7829,0.7579,0.1608); rgb(192pt)=(0.7945,0.7554,0.157); rgb(193pt)=(0.806,0.7529,0.1546); rgb(194pt)=(0.8172,0.7505,0.1535); rgb(195pt)=(0.8281,0.7481,0.1536); rgb(196pt)=(0.8389,0.7457,0.1546); rgb(197pt)=(0.8495,0.7435,0.1564); rgb(198pt)=(0.86,0.7413,0.1587); rgb(199pt)=(0.8703,0.7392,0.1615); rgb(200pt)=(0.8804,0.7372,0.165); rgb(201pt)=(0.8903,0.7353,0.1695); rgb(202pt)=(0.9,0.7336,0.1749); rgb(203pt)=(0.9093,0.7321,0.1815); rgb(204pt)=(0.9184,0.7308,0.189); rgb(205pt)=(0.9272,0.7298,0.1973); rgb(206pt)=(0.9357,0.729,0.2061); rgb(207pt)=(0.944,0.7285,0.2151); rgb(208pt)=(0.9523,0.7284,0.2237); rgb(209pt)=(0.9606,0.7285,0.2312); rgb(210pt)=(0.9689,0.7292,0.2373); rgb(211pt)=(0.977,0.7304,0.2418); rgb(212pt)=(0.9842,0.733,0.2446); rgb(213pt)=(0.99,0.7365,0.2429); rgb(214pt)=(0.9946,0.7407,0.2394); rgb(215pt)=(0.9966,0.7458,0.2351); rgb(216pt)=(0.9971,0.7513,0.2309); rgb(217pt)=(0.9972,0.7569,0.2267); rgb(218pt)=(0.9971,0.7626,0.2224); rgb(219pt)=(0.9969,0.7683,0.2181); rgb(220pt)=(0.9966,0.774,0.2138); rgb(221pt)=(0.9962,0.7798,0.2095); rgb(222pt)=(0.9957,0.7856,0.2053); rgb(223pt)=(0.9949,0.7915,0.2012); rgb(224pt)=(0.9938,0.7974,0.1974); rgb(225pt)=(0.9923,0.8034,0.1939); rgb(226pt)=(0.9906,0.8095,0.1906); rgb(227pt)=(0.9885,0.8156,0.1875); rgb(228pt)=(0.9861,0.8218,0.1846); rgb(229pt)=(0.9835,0.828,0.1817); rgb(230pt)=(0.9807,0.8342,0.1787); rgb(231pt)=(0.9778,0.8404,0.1757); rgb(232pt)=(0.9748,0.8467,0.1726); rgb(233pt)=(0.972,0.8529,0.1695); rgb(234pt)=(0.9694,0.8591,0.1665); rgb(235pt)=(0.9671,0.8654,0.1636); rgb(236pt)=(0.9651,0.8716,0.1608); rgb(237pt)=(0.9634,0.8778,0.1582); rgb(238pt)=(0.9619,0.884,0.1557); rgb(239pt)=(0.9608,0.8902,0.1532); rgb(240pt)=(0.9601,0.8963,0.1507); rgb(241pt)=(0.9596,0.9023,0.148); rgb(242pt)=(0.9595,0.9084,0.145); rgb(243pt)=(0.9597,0.9143,0.1418); rgb(244pt)=(0.9601,0.9203,0.1382); rgb(245pt)=(0.9608,0.9262,0.1344); rgb(246pt)=(0.9618,0.932,0.1304); rgb(247pt)=(0.9629,0.9379,0.1261); rgb(248pt)=(0.9642,0.9437,0.1216); rgb(249pt)=(0.9657,0.9494,0.1168); rgb(250pt)=(0.9674,0.9552,0.1116); rgb(251pt)=(0.9692,0.9609,0.1061); rgb(252pt)=(0.9711,0.9667,0.1001); rgb(253pt)=(0.973,0.9724,0.0938); rgb(254pt)=(0.9749,0.9782,0.0872); rgb(255pt)=(0.9769,0.9839,0.0805)},
]
\addplot [forget plot] graphics [xmin=-0.390625, xmax=100.390625, ymin=-50.87890625, ymax=26100.87890625] {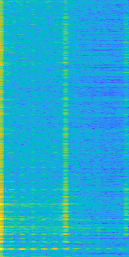};
\draw[color=red,dashed,ultra thick]  (0.5,200) rectangle (80,260); 
\end{axis}

\end{tikzpicture}%

%% file: figures/measurement_setup.tex
\usetikzlibrary{shapes.geometric,calc}

\begin{tikzpicture}[scale=0.8, every node/.style={scale=0.8}]
\def\nodewidth{2.5cm}
\def\nodeheight{2.25cm}
\newcommand{\AxisRotator}[1][rotate=0]{%
    \tikz [x=0.25cm,y=0.66cm,line width=.2ex,-stealth,#1] \draw (0,0) arc (-140:150:1 and 1);%
}
\coordinate (engine_center) at (-2,-2);
\coordinate (gear_center) at ($ (engine_center) + (4,0)$);
\coordinate (converter_center) at (0.0,1);

\node (engine) [rounded corners,fill=gray!20,draw,shape aspect=2.5,minimum height=\nodeheight,minimum width=\nodeheight,label={[rotate=0]center: Engine}] at (engine_center) {} ;
\draw[line width = 1pt] (engine.south) -- ($(engine.south) + (0.5,-1)$);
\draw[line width = 1pt] (engine.south) -- ($(engine.south) + (-0.5,-1)$);

\draw[ultra thick] ($(engine.north east) + (-0.25,0)$) -- node[pos = 1.6,label={[xshift=0.5cm, yshift=0.0cm]Sensor A}]{} ($(engine.north east) + (-0.25,0) + (0,0.25)$);

\draw[ultra thick] ($(engine.north west) + (0.25,0)$) -- node[pos = 1.6, label={[xshift=-0.5cm, yshift=0.0cm]Sensor B}]{} ($(engine.north west) + (0.25,0) + (0,0.25)$);

\node (gear) [rounded corners,fill=gray!20,draw,align = center,minimum height=\nodeheight,minimum width=\nodewidth,label={[rotate=0]center: }] at (gear_center) {Load \\ machine};

\draw[line width = 1pt] (gear.south) -- ($(gear.south) + (0.5,-1)$);
\draw[line width = 1pt] (gear.south) -- ($(gear.south) + (-0.5,-1)$);

\draw[ultra thick,line width=12pt, draw= black ] (engine) --  node [pos=0.21] {\AxisRotator[rotate=0]} node[pos=0.5, label={[align=center]below:{Shaft/\\Coupling}}]{} (gear);

\node[draw, align = center,rounded corners,fill=gray!20] (converter) at (converter_center) {Traction \\ converter}; 

\node[fill=black,draw,minimum width = 0.5cm, minimum height = 0.25cm] (box_engine) at ($(engine.north) + (0,0.13)$){};

\node[fill=black,draw,minimum width = 0.5cm, minimum height = 0.25cm] (box_load) at ($(gear.north) + (0,0.13)$){};
\draw[thick] ($(converter.west) + (0,0.15)$) to[bend right]  ($(box_engine.north west) + (0.1,0)$);
\draw[thick] (converter.west) to [bend right] (box_engine);
\draw[thick] ($(converter.west) + (0,-0.15)$) to[bend right] ($(box_engine.north east) + (-0.1,0)$);

\draw[thick] ($(converter.east) + (0,-0.15)$) to[bend left]  ($(box_load.north west) + (0.1,0)$);
\draw[thick] (converter.east) to [bend left] (box_load);
\draw[thick] ($(converter.east) + (0,0.15)$) to[bend left]  ($(box_load.north east) + (-0.1,0)$);
\node (n)[cylinder, fill=gray!30,thick,aspect=0.5,minimum height=0.4cm,minimum width=1cm,shape border      
   rotate=0] at ($(engine.center) +(1.9,0)$) {};

\end{tikzpicture}

%% file: figures/confusion_matrix.tex
\begin{tikzpicture}[scale=1.15,every node/.style={scale=1.15}]

\node[draw,fill=green,minimum width = 1.25cm] (tp) at (0,0) {TP};
\node[draw,fill=red,minimum width = 1.25cm] (fp) at (1.6,0) {FP};
\node[draw,fill=red,minimum width = 1.25cm] (tp) at (0,-0.7) {FN};
\node[draw,fill=green,minimum width = 1.25cm] (fp) at (1.6,-0.7) {TN};

\node[rotate=90] (pred) at (-1.4,-0.5) {\scriptsize Predicted Class};
\node[rotate=90] (pred) at (-1,-0.8) {\scriptsize N};
\node[rotate=90] (pred) at (-1,0) {\scriptsize P};
\node[rotate=0] (pred) at (0,-1.35) {\scriptsize P};
\node[rotate=0] (pred) at (1.6,-1.35) {\scriptsize N};


\node[] (truth) at (0.8,-1.8) {\scriptsize True Class};

\draw[ thick] (-0.8,-0.35) -- (2.4,-0.35);
\draw[ thick] (0.8,-1.1) -- (0.8,0.35);
\draw[ thick] (-0.8,-1.1) -- (2.4,-1.1);
\draw[ thick] (-0.8,0.35) -- (2.4,0.35);
\draw[ thick] (-0.8,-1.1) -- (-0.8,0.35);
\draw[ thick] (2.4,-1.1) -- (2.4,0.35);
\end{tikzpicture}

%% file: figures/mfcc_sensorB_only_healthy.tex
%
%
\definecolor{mycolor1}{rgb}{0.00000,0.44700,0.74100}%
\definecolor{mycolor2}{rgb}{0.85000,0.32500,0.09800}%
\definecolor{mycolor3}{rgb}{0.92900,0.69400,0.12500}%
\definecolor{mycolor4}{rgb}{0.49400,0.18400,0.55600}%
\begin{tikzpicture}

\begin{axis}[%
width=0.951\fwidth,
height=\fheight,
at={(0\fwidth,0\fheight)},
scale only axis,
xmin=0,
xmax=40,
ymin=0,
ymax=100,
ylabel = $\%$,
xlabel = {Number of MFCCs},
xmajorgrids,
ymajorgrids,
axis background/.style={fill=white},
legend style={legend cell align=left, align=left, draw=white!15!black, legend pos = outer north east}
]
\addplot [color=mycolor1, mark=o, mark options={solid, mycolor1}]
  table[row sep=crcr]{%
1	35.13\\
2	53.33\\
3	64\\
4	82.1\\
5	82.39\\
6	87.94\\
7	66.12\\
8	72.08\\
9	90.69\\
10	91.31\\
11	91.91\\
12	90.45\\
13	84.18\\
14	86.73\\
15	86.72\\
16	88.52\\
17	80.76\\
18	81.64\\
19	81.64\\
20	82.47\\
21	83.36\\
22	82.97\\
23	98.98\\
24	98.73\\
25	98.78\\
26	98.88\\
27	98.73\\
28	98.83\\
29	98.84\\
30	98.9\\
31	98.7\\
32	98.73\\
33	98.76\\
34	98.81\\
35	98.68\\
36	98.58\\
37	98.53\\
38	97.05\\
39	98.81\\
40	98.87\\
};
\addlegendentry{Acc}

\addplot [color=mycolor2, mark=square, mark options={solid, mycolor2}]
  table[row sep=crcr]{%
1	0.439999999999998\\
2	0.760000000000005\\
3	1.39\\
4	1.61\\
5	0.890000000000001\\
6	0.88000000000001\\
7	0.429999999999993\\
8	0.480000000000004\\
9	1.57000000000001\\
10	1.75999999999999\\
11	1.75999999999999\\
12	1.39\\
13	0.939999999999998\\
14	1\\
15	0.980000000000004\\
16	1.09\\
17	0.790000000000006\\
18	0.799999999999997\\
19	0.790000000000006\\
20	0.799999999999997\\
21	0.849999999999994\\
22	0.900000000000006\\
23	2.38000000000001\\
24	2.29000000000001\\
25	2.28\\
26	2.37\\
27	2.21000000000001\\
28	2.37\\
29	2.49000000000001\\
30	2.63\\
31	2.56\\
32	2.60000000000001\\
33	2.62\\
34	2.64\\
35	2.68000000000001\\
36	2.56999999999999\\
37	2.53\\
38	2.14999999999999\\
39	3.18000000000001\\
40	3.3\\
};
\addlegendentry{FPR}

\addplot [color=mycolor3, mark=triangle, mark options={solid, mycolor3}]
  table[row sep=crcr]{%
1	76.03\\
2	54.63\\
3	41.99\\
4	20.72\\
5	20.5\\
6	13.99\\
7	39.68\\
8	32.68\\
9	10.65\\
10	9.89\\
11	9.17999999999999\\
12	10.97\\
13	18.4\\
14	15.4\\
15	15.42\\
16	13.27\\
17	22.44\\
18	21.4\\
19	21.4\\
20	20.43\\
21	19.38\\
22	19.82\\
23	0.780000000000001\\
24	1.09\\
25	1.03\\
26	0.900000000000006\\
27	1.11\\
28	0.969999999999999\\
29	0.929999999999993\\
30	0.829999999999998\\
31	1.08\\
32	1.03999999999999\\
33	1\\
34	0.939999999999998\\
35	1.08\\
36	1.22\\
37	1.29000000000001\\
38	3.09\\
39	0.849999999999994\\
40	0.75\\
};
\addlegendentry{FNR}

\addplot [color=mycolor4, mark=diamond, mark options={solid, mycolor4}]
  table[row sep=crcr]{%
1	61.76\\
2	72.31\\
3	78.31\\
4	88.84\\
5	89.3\\
6	92.56\\
7	79.95\\
8	83.42\\
9	93.89\\
10	94.17\\
11	94.53\\
12	93.82\\
13	90.33\\
14	91.8\\
15	91.8\\
16	92.82\\
17	88.38\\
18	88.9\\
19	88.9\\
20	89.38\\
21	89.89\\
22	89.64\\
23	98.42\\
24	98.31\\
25	98.34\\
26	98.36\\
27	98.34\\
28	98.33\\
29	98.29\\
30	98.27\\
31	98.18\\
32	98.18\\
33	98.19\\
34	98.21\\
35	98.12\\
36	98.11\\
37	98.09\\
38	97.38\\
39	97.98\\
40	97.97\\
};
\addlegendentry{BA}
\end{axis}
\end{tikzpicture}%

%% file: IN21_novel_features_bearing_fault.bbl
\begin{thebibliography}{10}

\bibitem{randall_rb_rolling_2011}
{Randall, R.B.} and {Antoni, J.}
\newblock Rolling element bearing diagnostics—a tutorial.
\newblock {\em Mechanical Systems and Signal Processing}, 25(2):485--520, 2011.

\bibitem{zhang2017a}
{Zhang, Z.}, {Entezami, M.}, {Stewart, E.}, and {Roberts, C.}
\newblock Enhanced fault diagnosis of roller bearing elements using a
  combination of empirical mode decomposition and minimum entropy
  deconvolution.
\newblock {\em Proc. Institution of Mechanical Engineers, Part C: Journal of
  Mechanical Engineering Science}, 231(4):655--671, 2017.

\bibitem{wei2019}
{Wei, Y.}, {Li, Y.}, {Xu, M.}, and {Huang, W.}
\newblock A review of early fault diagnosis approaches and their applications
  in rotating machinery.
\newblock {\em Entropy}, 21(4):409--435, Apr 2019.

\bibitem{Elasha2014}
{Elasha, F.}, {Ruiz-Carcel, C.}, {Mba, D.}, and {Chandra, P.}
\newblock A comparative study of the effectiveness of adaptive filter
  algorithms, spectral kurtosis and linear prediction in detection of a
  naturally degraded bearing in a gearbox.
\newblock {\em Journal of Failure Analysis and Prevention}, 14(5):623--636, Jan
  2014.

\bibitem{hamadache2019}
{Hamadache, M.}, {Jung, J. H.}, {Park, J.}, and {Youn, B. D.}
\newblock A comprehensive review of artificial intelligence-based approaches
  for rolling element bearing {PHM}: shallow and deep learning.
\newblock {\em JMST Advances}, pages 1--27, 2019.

\bibitem{neupane2020}
{Neupane, D.} and {Seok, J.}
\newblock {Bearing} {Fault} {Detection} and {Diagnosis} using {Case} {Western}
  {Reserve} {University} {Dataset} with {Deep} {Learning} {Approaches}: A
  {Review}.
\newblock {\em IEEE Access}, pages 1--26, 2020.

\bibitem{oppenheim}
{Oppenheim, A. V.} and {Schafer, R. W.}
\newblock {\em Discrete-Time Signal Processing}.
\newblock Prentice Hall Press, USA, 3rd edition, 2009.

\bibitem{vakharia_comparison_2016}
{Vakharia, V.}, {Gupta, V. K.}, and {Kankar, P. K.}
\newblock A comparison of feature ranking techniques for fault diagnosis of
  ball bearing.
\newblock {\em Soft Computing}, 20(4):1601--1619, April 2016.

\bibitem{vargas_2020}
{Vargas-Machuca, J.}, {Garcia, F.}, and {Coronado, A. M.}
\newblock Detailed {Comparison} of {Methods} for {Classifying} {Bearing}
  {Failures} {Using} {Noisy} {Measurements}.
\newblock {\em Journal of Failure Analysis and Prevention}, 20(3):744--754,
  June 2020.

\bibitem{nayana_analysis_2017}
{Nayana, B. R.} and {Geethanjali, P.}
\newblock Analysis of {Statistical} {Time}-{Domain} {Features} {Effectiveness}
  in {Identification} of {Bearing} {Faults} {From} {Vibration} {Signal}.
\newblock {\em IEEE Sensors Journal}, 17(17):5618--5625, September 2017.

\bibitem{br_feature_2019}
{BR, N.} and {Geethanjali, P.}
\newblock {Feature Extraction} for {Bearing} {Fault} {Diagnosis} in {Noisy}
  {Environment}: {A} {Study}.
\newblock In {\em 2019 {Innovations} in {Power} and {Advanced} {Computing}
  {Technologies} (i-{PACT})}, volume~1, pages 1--5, March 2019.

\bibitem{arun2018}
{Arun, P.}, {Abraham Lincon, S.}, and {Prabhakaran, N.}
\newblock Detection and characterization of bearing faults from the frequency
  domain features of vibration.
\newblock {\em IETE Journal of Research}, 64(5):634--647, 2018.

\bibitem{yuan_fault_2020}
{Yuan, Y.}, {Chen, C.}, and {Yuan, Y.}
\newblock Fault detection of rolling bearing based on principal component
  analysis and empirical mode decomposition.
\newblock {\em AIMS Mathematics}, 5(6):5916--5938, 2020.

\bibitem{schafer2007}
{Lawrence, R. W.} and {Schafer, R. W.}
\newblock {\em Introduction to Digital Speech Processing}.
\newblock Now Publishers Inc., MA, USA, 2007.

\bibitem{Kollmeier1994}
{Kollmeier, B.} and {Koch, R.}
\newblock Speech enhancement based on physiological and psychoacoustical models
  of modulation perception and binaural interaction.
\newblock {\em The Journal of the Acoustical Society of America},
  95(3):1593—1602, March 1994.

\bibitem{Tchorz2018}
{Tchorz, J.}
\newblock Combination of amplitude modulation spectrogram features and {MFCCs}
  for acoustic scene classification.
\newblock Technical report, DCASE2018 Challenge, September 2018.

\bibitem{schoelkopf2001}
{Schölkopf, B.}, {Platt, J. C.}, {Shawe-Taylor, J.}, {Smola, A. J.}, and
  {Williamson, R. C.}
\newblock Estimating the support of a high-dimensional distribution.
\newblock {\em Neural Computation}, 13(7):1443--1471, 2001.

\bibitem{bishop2006}
{Bishop, C. M.}
\newblock {\em Pattern Recognition and Machine Learning}.
\newblock Springer, Berlin, DE, 2006.

\end{thebibliography}
